\documentclass[12pt]{iopart}

\expandafter\let\csname equation*\endcsname=\relax
\expandafter\let\csname endequation*\endcsname=\relax

\usepackage{amsmath}%
\usepackage{amssymb}%
\usepackage{amsthm}%
\usepackage{graphicx}%
\usepackage{dsfont}%

\newcommand{\pa}[1]{\left( #1 \right)}
\newcommand{\ac}[1]{\left\{ #1 \right\}}
\newcommand{\crochet}[1]{\left[ #1 \right]}

\newcommand{\D}{\Delta}

\newcommand{\p}{\rho}
\newcommand{\s}{\sigma}

\newcommand{\q}{\theta}

\newcommand{\NN}{\mathds{N}}

\newcommand{\RR}{\mathds{R}}

\newcommand{\ie}{\textit{i.e.}\ }

\newcommand{\del}{\partial}

\newenvironment{aleq}{\begin{equation}\begin{aligned}}{\end{aligned}\end{equation}}
\newenvironment{aleq*}{\begin{equation*}\begin{aligned}}{\end{aligned}\end{equation*}}
\newenvironment{gaeq}{\begin{equation}\begin{gathered}}{\end{gathered}\end{equation}}
\newenvironment{gaeq*}{\begin{equation*}\begin{gathered}}{\end{gathered}\end{equation*}}
\newenvironment{eqe}{\begin{equation}}{\end{equation}}

\begin{document}
\title{Symmetry groups of non-stationary planar ideal plasticity}
\author{Vincent Lamothe\\D\'epartement de math\'ematiques et statistiques,\\Universit\'e de Montr\'eal, C.P. 6128, Succ. Centre-ville,\\
 Montr\'eal, (QC) H3C 3J7, Canada\ead{lamothe@crm.umontreal.ca}}

\begin{abstract}
This paper is a study of the Lie groups of point symmetries admitted by a system describing a non-stationary planar flow of an ideal plastic material. For several types of forces involved in the system, the infinitesimal generators which generate the Lie algebra of symmetries have been obtained. In the case of a monogenic force, the classification of one- and two- dimensional subalgebras into conjugacy classes under the action of the group of automorphisms has been accomplished. The method of symmetry reduction is applied for certain subalgebra classes in order to obtain invariant solutions.
\end{abstract}
\ 
\ \\
\submitto{J. Phys. A}
\maketitle%
\section{Introduction}
This paper is a study from the point of view of the symmetry group of a
system of partial differential equations (PDEs) modeling a planar
flow of an ideal plastic material  which is assumed to be
incompressible and subject to a force $F=(F_1,F_2)$, where $F_1$ and
$F_2$ are the components along the $x$ and $y$ axes respectively. The force
is a function of the independent variables $(t,x,y)$ and of the
velocity components $(u,v)$ relative to the $x$ and $y$ axes
respectively. The considered system depends on the four dependent
variables $\sigma$, $\theta$, $u$ et $v$, the significance of which
is given below. It is composed of the following four quasilinear
PDEs \cite{Kat:1,Hill, Chak}
\begin{aleq}\label{eq:1}
&(a)\quad &&\s_x-\pa{\q_x \cos 2\q+\q_y \sin2\q}+\p \pa{F_1-u_t -u u_x -v u_y}=0,\\
&(b) &&\s_y-\pa{\q_x\sin2\q - \q_y \cos 2\q}+\p \pa{F_2-v_t - u v_x - v v_y}=0,\\
&(c) &&(u_y+v_x)\sin 2\q + (u_x-v_y)\cos2\q=0,\\
&(d) &&u_x+v_y=0,
\end{aleq}%
where we used the notation $\s_x=\del \sigma/\del
x$, $\s_y=\del \sigma/\del y$, \textit{etc}. The quantity $\p=(2k)^{-1}\tilde{\p}$ is constant and represents the density $\tilde{\p}$ of the material divided by twice the yield limit $k$. If we denote the mean pressure by $p$, then the dependent variables $\sigma=(2 k)^{-1} p$ and the angle $\theta$ define the strain tensor. Equations (\ref{eq:1}.a) and (\ref{eq:1}.b) are the differential equations of motion of continuum mechanics for the planar problem. The equations of Saint-Venant-Von Mises plasticity theory are considered in this paper. In the case of a planar flow, they reduce to the single equation (\ref{eq:1}.c) under the assumption of incompressibility required through (\ref{eq:1}.d).
\paragraph{}In order to find new solutions of the system consisting of the two equations (\ref{eq:1}.a) and (\ref{eq:1}.b) with $\rho=0$ (the statically determined problem), \cite{Senashov:2007,Senashov:2009} acted with transformations of the symmetry group on known solutions of problems with particular boundary conditions, \ie the Nada\"{\i}'s solution \cite{Nadai:circularSol} for a circular cavity subject to a constraint and a normal shear at the surface, and also on the Prandtl solution \cite{Prandtl:solPlas} for a block compressed between two plates. The stationary version ($\rho=0$) of the system (\ref{eq:1}) has also been studied \cite{Czyz:1974}. Solutions were found involving simple and double Riemann waves by using the method of characteristics. However, as is often the case with this method, solutions rely on numerical integration in order to obtain velocity components $u$ and $v$. A study of the system (\ref{eq:1}) from the group-theoretical point of view has been done for the stationary case \cite{Lamothe:JMP:2012,Lamothe:JPA:2012}. Such a study has never been carried out for the non-stationary case and for different types of force involved in the system.
\paragraph{}The purpose of this paper is to make a systematic study of system (\ref{eq:1}) from the point of view of the Lie group of symmetry transformations of this system in order to obtain new analytic solutions. More specifically, the infinitesimal generators of symmetries which generate the Lie algebra $\mathcal{L}$ associated with the group $G$ are obtained. Different algebras are obtained depending on the force $(F_1,F_2)$  chosen. Next, for a monogenic force (which admits the largest possible symmetry group), a classification of the one- and two- dimensional subalgebras into conjugacy classes under the action of the group of automorphisms of $G$ is performed, based on the techniques introduced in \cite{PateraWinter:1,PateraWinter:2}. This allows one to obtain invariant and partially invariant (with defect structure $\delta=1$ in the sense of Ovsiannikov \cite{Ovsiannikov:Group_Analysis}) solutions. The paper is organized as follows. The symmetry generators, commutation relations and preliminary  classification of forces are presented in Section \ref{sec:2}. Section \ref{sec:3} is concerned with the classification of subalgebras into conjugacy classes associated with the symmetry group of system (\ref{eq:1}) in the case where the intervening force is monogenic. Symmetry reductions corresponding to certain interesting subalgebras are performed in Section \ref{sec:4}. Section \ref{sec:con} contains final remarks and possible future developments.
\section{Algebras of symmetries}\label{sec:2}
Since a force $F$ is involved in the PDEs system (\ref{eq:1}), the application of the classical infinitesimal techniques, as presented in \cite{Olver:Application_of_Lie}, results in different symmetry groups depending on the type of forces considered. This section is concerned with the problem of classification of admissible types of force and their associated symmetry groups.
\paragraph{}First, we apply the symmetry criterion (see \textit{e.g.} \cite{Olver:Application_of_Lie}) to the system (\ref{eq:1}) assuming that the force $F$ is a function of the independent variables $t,x,y$ and the velocity components $u$ and $v$. The dependency of the force $F$ on $u$ and $v$ allows us to consider friction phenomena. Consider a vector field of the form
\begin{eqe}\label{eq:2}
X=\xi_1\del_x+\xi_2\del_y+\xi_3\del_t+\phi_1\del_u+\phi_2\del_v+\phi_3\del_\sigma+\phi_4\del_\theta,
\end{eqe}%
where the coefficients $\xi_i$, $\phi_j$, $i=1,2,3$, $j=1,\ldots,4$ are functions of the independent variables $t,x,y$ and the dependent variables $u,v,\sigma,\theta$. The symmetry criterion applied to the system (\ref{eq:1}) provides more than one hundred linear first-order PDEs for the coefficients $\xi_i$ and $\phi_j$ of the vector field (\ref{eq:2}), the so-called determining equations, which are omitted here so not as to lengthen this paper. Taking aside two of these determining equations, given below in (\ref{eq:4}), the most general solution of all the others determining equations for an arbitrary force is found to be
\begin{gaeq}\label{eq:3}
\xi_0=c_1t+c_0,\quad \xi_1=c_1 x+\tau_1(t)y+\tau_2(t),\quad \xi_2=c_1y-\tau_1(t)x+\tau_3(t),\\
\phi_1=\tau_1(t)v+\dot{\tau}_1(t)y+\dot{\tau}_2(t),\quad \phi_2=-\tau_1(t)v-\dot{\tau}_1(t)x+\dot\tau_3(t),\\
\phi_3=\rho\eta(t,x,y),\quad\phi_4=-\tau_1(t),
\end{gaeq}%
where $c_0$ and $c_1$ are real parameters, $\tau_1$, $\tau_2$ and $\tau_3$ are arbitrary functions of the time $t$, the arbitrary function $\eta$ depend on $t,x,y$ and the dot over a function denotes the time derivative. The two remaining determining equations
\begin{aleq}\label{eq:4}
&(a)\qquad
&&\eta_x(t,x,y)+c_1\pa{F_1+tF_{1,t}+xF_{1,x}+yF_{1,y}}+c_0F_{1,t}-\tau_1''(t)y\\
&
&&+\tau_1'(t)\pa{yF_{1,u}-xF_{1,v}-2v}+\tau_1(t)\pa{yF_{1,x}-xF_{1,y}+vF_{1,u}-uF_{1,v}-F_2}\\
&
&&+\tau_2(t)F_{1,x}+\tau_2'(t)F_{1,u}-\tau_2''(t)+\tau_3(t)F_{1,y}+\tau_3'(t)F_{1,v}=0,\\
&(b) &&\eta_y(t,x,y)+c_1\pa{F_2+tF_{2,t}+xF_{2,x}+yF_{2,y}}+c_0F_{2,t}+\tau_1''(t)x\\
&
&&+\tau_1'(t)\pa{yF_{2,u}-xF_{2,v}+2u}+\tau_1(t)\pa{yF_{2,x}-xF_{2,y}+vF_{2,u}-uF_{2,v}+F_1}\\
&
&&+\tau_2(t)F_{2,x}+\tau_2'(t)F_{2,u}+\tau_3(t)F_{2,y}+\tau_3'(t)F_{2,v}-\tau_3''(t)=0.
\end{aleq}%
have to be satisfied by a proper choice of the parameters $c_0$, $c_1$, the arbitrary functions $\eta$, $\tau_i$, $i=1,2,3,$ and the force components $F_1$ and $F_2$.
\paragraph{}If one is interested by the symmetry algebra admitted by the system (\ref{eq:1}) for any kind of forces, then equations (\ref{eq:4}) have to be solved for $c_0$, $c_1$, $\eta$ and $\tau_i$, $i=1,2,3$, in such a manner that the force components $F_1$ and $F_2$ stay arbitrary. So, the coefficients of the force components $F_1$ and $F_2$ as well as the coefficients of all their partial derivatives must vanish. It is easy to verify that the only possible solution in this case is
\begin{eqe}\label{eq:5}
c_0=c_2=c_2=\tau_1=\tau_2=\tau_3=0\quad\text{and}\quad \eta=h(t),
\end{eqe}%
where $h(t)$ is an arbitrary function of time. Consequently, the infinite dimensional Lie algebra spanned by the vector fields of the form
\begin{eqe}\label{eq:6}
S_{h(t)}=h(t)\del_\sigma,
\end{eqe}%
is a symmetry algebra admitted by the system (\ref{eq:1}) for any type of force. A vector field of the form (\ref{eq:6}) is an infinitesimal generator of a transformation that consists of adding an arbitrary function of time to the dependent variable $\sigma$ and leaving unchanged the other variables involved in the system.
\paragraph{}Together with the vector field (\ref{eq:6}), more vector fields can be included in the symmetry algebra of the system (\ref{eq:1}) for more specific types of force. It should be noted that, if the PDEs (\ref{eq:4}) are solved in such a manner that the force components $F_1$ and $F_2$ involve the parameters $c_0$, $c_1$ and the functions $\tau_i$ and $\eta$, then these parameters and functions must be considered as parameterizing the force $F$ and will no longer be available to span the Lie algebra of symmetry. Consequently, in order to obtain a symmetry algebra of large dimension, one has to ensure that the force components $F_1$ and $F_2$ involve the lowest possible number of parameters $c_0$, $c_1$ and parameterizing functions $\eta$ and $\tau_i$, $i=1,2,3$, in the solution of the PDEs (\ref{eq:4}).
\paragraph{}Since the derivatives of $\eta$ with respect to $x$ and $y$ appear in the equations (\ref{eq:4}.a) and (\ref{eq:4}.b) respectively, the compatibility of the mixed derivatives must be satisfied. This compatibility condition can be written as
\begin{gaeq}\label{eq:7}
c_0G_t+c_1\pa{2G+tG_t+xG_x+yG_y}+\tau_1\pa{yG_x-xG_y+vG_u-uG_v}\\
+\dot{\tau}_1\pa{yG_u-xG_v+F_{1,u}+F_{2,v}}-2\ddot{\tau}_1 +\tau_2G_x+\dot{\tau}_2G_u+\tau_3G_y+\dot{\tau}_3G_v=0,
\end{gaeq}%
where $G=F_{1,y}-F_{2,x}$. If this compatibility condition is satisfied, then the equations (\ref{eq:4}) can be integrated by quadrature for $\eta(t,x,y)$ provided that the derivatives with respect to $u$ and $v$ of equations (\ref{eq:4}) cancel out. Indeed, since $\eta(t,x,y)$ does not depend on $u$ and $v$, the derivatives $\eta_u$ and $\eta_v$ vanish, which leads us to four equations obtained as differential consequences of the PDEs (\ref{eq:4}). In order to obtain a large group of symmetries, we assume that the coefficients in front of the constants $c_0$ and $c_1$, together with the coefficients in front of the arbitrary functions $\tau_2(t)$, $\tau_3(t)$ and their derivatives, vanish. This implies that
 \begin{eqe}\label{eq:p:1}
 G=F_{1,y}-F_{2,x}=0,
 \end{eqe}%
 and therefore equation (\ref{eq:7}) reduces to
 \begin{eqe}\label{eq:p:2}
 -\dot{\tau}_1(t)\pa{F_{1,u}+F_{2,v}}+2\ddot{\tau}_1(t)=0.
 \end{eqe}%
 The general solution of equation (\ref{eq:p:1}) takes the form
 \begin{eqe}\label{eq:p:3}
 F_1=U_x(t,x,y,u,v),\quad F_2=U_y(t,x,y,u,v),
 \end{eqe}%
where $U$ is an arbitrary differentiable function. If we introduce the force of the
form (\ref{eq:p:3}) into equation (\ref{eq:p:2}) and into the
differential consequences with respect to $u$ and $v$ of
equations (\ref{eq:4}), and imposing the condition that the
coefficients $c_0$, $c_1$, $\tau_2(t)$ and $\tau_3(t)$ all vanish,
we find that the components of the force do not depend on $u$ and $v$.
In this case, equation (\ref{eq:p:2}) implies that $\tau_1(t)$ is a
constant, which we denote $c_2$. The force (\ref{eq:p:3}) then
reduces to a monogenic type, \ie
\begin{eqe}\label{eq:p:4}
F_1=V_x(t,x,y),\quad F_2=V_y(t,x,y),
\end{eqe}%
where $V(t,x,y)$ is an arbitrary function. Therefore, we can solve equations
(\ref{eq:4}) for the quantity $\eta$, in terms of the parameters
$c_0$, $c_1$, $c_2$ and of the functions $\tau_2$, $\tau_3$, which takes
the form
\begin{aleq*}
\eta=&-c_1\pa{tV_t+xV_x+yV_y}-c_2V_t+c_3\pa{xV_y-yV_x}-\tau_2(t)V_x+\tau_3(t)V_y\\
&+\tau_2''(t)x+\tau_3''(t)y+h(t).
\end{aleq*}%
The symmetry algebra associated with the force (\ref{eq:p:4}) is generated by the vector fields (\ref{eq:6}) and the following
\begin{gaeq}\label{eq:8}
P_0=\del_t-\rho V_t\del_\sigma,\\
B_x=\tau_2(t)\del_x+\tau_2'(t)\del_u-\rho\pa{\tau_2(t)
V_x-x\tau_2''(t)}\del_\sigma,\\
B_y=\tau_3(t)\del_y+\tau_3'(t)\del_v-\rho\pa{\tau_3(t)
V_y-y\tau_3''(t)}\del_\sigma,\\
D=t\del_t+x\del_x+y\del_y-\rho\pa{tV_t+xV_x+yV_y}\del_\sigma,\\
L=y\del_x-x\del_y+v\del_u-u\del_v+\rho\pa{xV_y-yV_x}\del_\sigma-\del_\theta,\\
\end{gaeq}%
where the functions $\tau_2(t)$, $\tau_3(t)$ and $h(t)$ are arbitrary. Due to
the arbitrariness of the functions $\tau_2(t)$, $\tau_3(t)$ and $h(t)$,
these vector fields generate an infinite-dimensional symmetry
algebra. The vector fields (\ref{eq:6}) and (\ref{eq:8}) are the infinitesimal generators of the symmetry group in the case of a monogenic force (\ref{eq:p:4}). Rather than make the group analysis of the Lie algebra spanned by the vector fields (\ref{eq:8}), one can equivalently consider the simpler case of the Lie algebra in the situation where no force is involved in the system (\ref{eq:1}). Indeed, the simple mapping
$$(t,x,y,u,v,\sigma,\theta)\mapsto(t,x,y,u,v,\sigma-\rho V(t,x,y),\theta),$$
where $V(t,x,y)$ is the function defining the force (\ref{eq:p:4}), maps solutions of the null force problem to solutions of the monogenic force problem. For this reason, in next section, a group analysis of the system (\ref{eq:1}) when $F_1=F_2=0$ will be done. Hence, we will be concerned with the Lie algebra spanned by (\ref{eq:6}) together with the vector fields
\begin{gaeq}\label{eq:9}
P_0=\del_t,\quad D=t\del_t+x\del_x+y\del_y,\quad L=y\del_x-x\del_y+v\del_u-u\del_v-\del_\theta,\\
X_{f(t)}=f(t)\del_x+\dot{f}(t)\del_u+\rho x\ddot{f}(t)\del_\sigma,\quad Y_{g(t)}=g(t)\del_y+\dot{g}(t)\del_v+\rho y\ddot{g}(t)\del_\sigma,
\end{gaeq}%
where $f(t)$ and $g(t)$ are arbitrary functions of time. The nonzero commutation relations are
\begin{gaeq}\label{eq:10}
\crochet{P_0,X_{f(t)}}=\dot{f}(t)\del_x+\ddot{f}(t)\del_u+\rho x\dddot{f}(t)\del_\sigma=X_{\dot{f}(t)},\\
\crochet{P_0,Y_{g(t)}}=\dot{g}(t)\del_y+\ddot{g}(t)\del_v+\rho y\dddot{g}(t)\del_\sigma=Y_{\dot{g}(t)},\\
\crochet{P_0,D}=P_0,\quad \crochet{P_0,S_{h(t)}}=\dot{h}(t)\del_\sigma=S_{\dot{h}(t)}\in\mathcal{S},\quad \crochet{D,S_{h(t)}}=t\dot{h}(t)\del_\sigma=S_{t\dot{h}(t)},\\
\crochet{D,X_{f(t)}}=\tilde{f}(t)\del_x+\dot{\tilde{f}}(t)\del_u+\rho x \ddot{\tilde{f}}(t)\del_\sigma=X_{\tilde{f}(t)}\quad\text{where }\tilde{f}(t)=t\dot{f}(t)-f(t),\\ \crochet{D,Y_{g(t)}}=\tilde{g}(t)\del_y+\dot{\tilde{g}}(t)\del_v+\rho x \ddot{\tilde{g}}(t)\del_\sigma=Y_{\tilde{g}(t)}\quad\text{where }\tilde{g}(t)=t\dot{g}(t)-g(t),\\
\crochet{L,X_{f(t)}}=f(t)\del_y+\dot{f}(t)\del_v+\rho y \ddot{f}(t)\del_\sigma=Y_{f(t)},\\
\crochet{L,Y_{g(t)}}=-\pa{g(t)\del_x+\dot{g}(t)\del_u+\rho x \ddot{g}(t)\del_\sigma}=-X_{g(t)}.
\end{gaeq}%
\subsection{Algebras of symmetries when the force depends on the velocity} \paragraph{}A complete classification of all possible types of force and their associated symmetry algebras will be performed in a future work. Nevertheless, some interesting types of forces depending on the velocity components are given. One interesting type of force which allows the consideration of friction or viscosity arises when we make the hypotheses $c_1=\kappa_1$, $\tau_1(t)=\kappa_2$, $\tau_2(t)=c_2$, $\tau_3(t)=c_3$ and $\eta(t,x,y)=s(t)$. In this case, the components of the force are given by
\begin{aleq}\label{eq:fr:1}
F_1&=\pa{uh_1(u^2+v^2)+vh_2(u^2+v^2)}\exp\pa{(\kappa_1/\kappa_2)\arctan(v/u)},\\
F_2&=\pa{vh_1(u^2+v^2)-uh_2(u^2+v^2)}\exp\pa{(\kappa_1/\kappa_2)\arctan(v/u)},
\end{aleq}%
where $h_1$ and $h_2$ are arbitrary functions of the velocity, while
$\kappa_1$ and $\kappa_2$ are the real parameters of the force. The
force components given in (\ref{eq:fr:1}) constitute a force with a
contribution along the velocity vector parameterized by the
arbitrary function $h_1$ and a contribution perpendicular to the
velocity vector parameterized by the function $h_2$. Depending on
the sign of the argument of the exponential and on the angle between
the velocity vector and the $u$-axis, the force is damped or
amplified. The symmetry algebra of the system (\ref{eq:1}) for a
force of type (\ref{eq:fr:1}) is generated by the generators
\begin{gaeq}\label{eq:fr:2}
P_0=\del_t,\quad P_1=\del_x,\quad P_2=\del_y,\quad P_\sigma=\rho s(t)\del_\sigma,\\
K=\kappa_1\pa{t\del_t+x\del_x+y\del_y}+\kappa_2\pa{y\del_x-x\del_y+v\del_u-u\del_v-\del_\theta}.
\end{gaeq}%
It is possible to add a contribution to the force which is a function of time, but only at the cost of losing the generator $P_0=\del_t$ in the basis of the symmetry algebra. This force takes the form
\begin{aleq}\label{eq:fr:3}
F_1&=\pa{uh_1(u^2+v^2)+vh_2(u^2+v^2)}\exp\pa{(\kappa_1/\kappa_2)\arctan(v/u)}\\
&+\pa{t+\kappa_0/\kappa_1}^{-1}\pa{\kappa_3\sin\pa{\frac{\kappa_2}{\kappa_1}\ln\pa{t+\frac{\kappa_0}{\kappa_1}}} +\kappa_4\cos\pa{\frac{\kappa_2}{\kappa_1}\ln\pa{t+\frac{\kappa_0}{\kappa_1}}}}\\
F_2&=\pa{vh_1(u^2+v^2)-uh_2(u^2+v^2)}\exp\pa{(\kappa_1/\kappa_2)\arctan(v/u)}\\
&+\pa{t+\kappa_0/\kappa_1}^{-1}\pa{-\kappa_3\cos\pa{\frac{\kappa_2}{\kappa_1}\ln\pa{t+\frac{\kappa_0}{\kappa_1}}} +\kappa_4\sin\pa{\frac{\kappa_2}{\kappa_1}\ln\pa{t+\frac{\kappa_0}{\kappa_1}}}},
\end{aleq}%
where $h_1$, $h_2$, are arbitrary functions and $\kappa_i$, $i=1,2,3,4$, are force parameters. The symmetry algebra is spanned by the generators
\begin{gaeq}\label{eq:fr:4}
P_1=\del_x,\quad P_2=\del_y,\quad P_\sigma=\rho s(t)\del_\sigma,\\
K=\kappa_0\del_t+\kappa_1\pa{t\del_t+x\del_x+y\del_y}+\kappa_2\pa{y\del_x-x\del_y+v\del_u-u\del_v-\del_\theta}.
\end{gaeq}%
A more extensive generalization of the force (\ref{eq:fr:1}) is possible, but it admits an algebra with a lower dimension. This force takes the form
\begin{aleq}\label{eq:fr:5}
F_1&=\pa{uh_1(u^2+v^2)+vh_2(u^2+v^2)}\exp\pa{(\kappa_1/\kappa_2)\arctan(v/u)}\\
&+t^{-1}\pa{x h_3\pa{\frac{x^2+y^2}{t^2}}+yh_4\pa{\frac{x^2+y^2}{t^2}}}\exp\pa{(\kappa_1/\kappa_2)\arctan\pa{y/x}}\\
F_1&=\pa{vh_1(u^2+v^2)-uh_2(u^2+v^2)}\exp\pa{(\kappa_1/\kappa_2)\arctan(v/u)}\\
&+t^{-1}\pa{y h_3\pa{\frac{x^2+y^2}{t^2}}-xh_4\pa{\frac{x^2+y^2}{t^2}}}\exp\pa{(\kappa_1/\kappa_2)\arctan\pa{y/x}},
\end{aleq}%
The associated symmetry algebra is generated by $P_0$ and $K$, as
defined in (\ref{eq:fr:2}). The force presents a contribution
parallel to the velocity vector corresponding to the terms
containing the arbitrary function $h_3$. It also has a contribution
perpendicular to the velocity vector corresponding to the terms
containing the function $h_4$. The forces presented above in this
subsection do not represent a complete classification of the forces
satisfying the PDEs (\ref{eq:4}). Nevertheless, they constitute interesting
examples of forces parameterized by real constants $\kappa_i$ and
arbitrary functions of one or two variables. The forces can be
expressed in terms of the components $u$ and $v$ of the velocity.
Consequently, they can be interpreted, for instance, as friction
phenomena. For some of the forces presented in this subsection, the
system (\ref{eq:1}) admits generators containing parameters of the
force in their expressions. This means that each force determined by
a particular choice of parameters $\kappa_i$ is associated with a set of generators specific to it.
\section{Subalgebra classification when no external force $F$ is involved}\label{sec:3}
A symmetry group of a PDE system is a Lie group of point transformations which maps solutions of the system to others solutions. So, from a particular solution of PDEs, one can generate a multi-parameter family of solutions using the symmetry group. Another interesting application of symmetry groups to PDEs is the symmetry reduction method (SRM) \cite{Olver:Application_of_Lie}. This method consist of looking for invariant solutions under the action of a specific subgroup of the symmetry group. The invariance requirement is imposed through side conditions established via the use of the vector fields that span the subgroup considered. A natural question that arises is the following. How two distinct solutions obtained by the use of the SRM, and corresponding to two distinct subgroups, can be related in the sense that one solution can be computed from the other by the application of a transformation of the symmetry group? The answer is that two invariant solutions can be calculated one from the other if their respective subgroups (under which they are invariant) are conjugated under the action of the symmetry group. Suppose that $G$ is a symmetry group and that $H_1$, $H_2\subset G$ are two distinct subgroups. This means that there exists $g\in G$ such that
$$gH_1g^{-1}=H_2.$$%
Hence, if one wants to find the set of all nonequivalent invariant solutions, in the previous sense, then he has to use the SRM once using a representative subgroup $H_i$ of each conjugacy classes $\bar{H}_i$, $\bar{H}_i=\ac{gH_ig^{-1}: g\in G}$. But it is well known that to each subgroup $H_i\subset G$ correspond a unique subalgebra $\mathcal{L}_i\subset\mathcal{L}$, where $\mathcal{L}$ is the Lie algebra associated to the Lie group $G$. Consequently, a classification in conjugacy classes under the action of $G$ consists in a list of representative subalgebras $\mathcal{L}_i\subset\mathcal{L}$, one for each class. The technique to achieve such a classification has been developed in \cite{PateraWinter:1}.
\paragraph{}The sequel of this Section will be concerned with the problem of classifying, into conjugacy classes, the subalgebras of the Lie algebra
\begin{eqe}\label{eq:SC:1}
\mathcal{L}=\langle L,D,P_0,X_{f(t)}, Y_{g(t)}, S_{h(t)}\rangle,
\end{eqe}%
spanned by the vector fields (\ref{eq:6}) and (\ref{eq:8}) which generate the symmetry group of the system (\ref{eq:1}) when no force is involved, that is when $F=(F_1,F_2)=0$. The angle bracket in the right hand side of (\ref{eq:SC:1}) is used to denote the Lie algebra spanned by the basis vector fields inside. It is convenient to decompose the Lie algebra (\ref{eq:SC:1}) into a semi-direct sum of the form
\begin{eqe}\label{eq:SC:2}
\mathcal{L}=\mathcal{F}\rhd\mathcal{S},
\end{eqe}%
where the infinite dimensional Abelian ideal $\mathcal{S}=\langle S_{h(t)}\rangle$ is spanned by the vector field (\ref{eq:6}) and the factor subalgebra $\mathcal{F}$ further decomposes into the semi-direct sum
\begin{eqe}\label{eq:SC:3}
\mathcal{F}=\mathcal{A}\rhd\mathcal{B},
\end{eqe}%
with
\begin{eqe}\label{eq:SC:4}
\mathcal{A}=\langle L,D,P_0\rangle\quad\text{and}\quad\mathcal{B}=\langle X_{f(t)},Y_{g(t)}\rangle.
\end{eqe}%
Such a decomposition allows us to work iteratively, first classifying the subalgebra $\mathcal{F}$ under the action of $\exp \mathcal{F}$ and next the Lie algebra $\mathcal{L}$ under the action of the whole symmetry group denoted $G=\exp\mathcal{L}$. One should note that $\mathcal{B}$ is Abelian and infinite dimensional. The procedure developed in \cite{PateraWinter:1,PateraWinter:2} for the classification of subalgebras, requires us to first classify the factor subalgebra (under its inner automorphisms) in the case of a semi-direct sum decomposition. So, the first step in classifying $\mathcal{F}$ is to classify the three-dimensional algebra $\mathcal{A}$ given by (\ref{eq:SC:4}). This has already been done in a work by P. Winternitz and J. Patera \cite{PateraWinter:1977}. The result of this classification is listed in Table \ref{tab:1}.
\begin{table}
\begin{center}
\begin{tabular}{clll}
  dimension & representative & normalizer group & hypotheses \\ \hline
   & $\langle L\rangle$ & $\exp\mathcal{A}$ &  \\
   & $\langle D\rangle$ & $\exp\langle L,D\rangle$ &  \\
  1 & $\langle P_0\rangle$ & $\exp\mathcal{A}$ &  \\
   & $\langle L+a D\rangle$ & $\exp\langle L,D\rangle$ & $a\in\RR, a\neq 0$ \\
   & $\langle L+P_0\rangle$ & $\exp\langle L,P_0\rangle$ &  \\ \hline
   & $\langle L,D\rangle$ & $\exp\mathcal{A}$ &  \\
  2 & $\langle L,P_0\rangle$ & $\exp\mathcal{A}$ &  \\
   & $\langle D+aL,P_0\rangle$ & $\exp\mathcal{A}$ & $a\in\RR$ \\ \hline
  3 & $\mathcal{A}$ & $\exp\mathcal{A}$ &
\end{tabular}
\end{center}
\caption{List of the representative subalgebras of $\mathcal{A}$}\label{tab:1}
\end{table}%
In order to complete the classification of $\mathcal{F}$, it remains to find representative subalgebras of $\mathcal{F}$ that cannot be conjugate to one of those listed in Table \ref{tab:1}. There exist two types of such subalgebras. The first type are the splitting subalgebras of $\mathcal{F}$ which consist of semi-direct sums of a factor subalgebra $\mathcal{A}_i\subset \mathcal{A}$ and an ideal subalgebra $\mathcal{B}_i\subset\mathcal{B}$. Since $\mathcal{L}$ is a Lie algebra and $\mathcal{B}\subset \mathcal{L}$, it suffices to check that a subspace $\mathcal{B}_j\subset\mathcal{B}$ satisfies $\crochet{\mathcal{B}_j,\mathcal{B}_j}\subset\mathcal{B}_j$ and $\crochet{\mathcal{A}_i,\mathcal{B}_j}\subset \mathcal{B}_j$ in order to ensure that $\mathcal{B}_j$ will form an ideal in the direct sum
\begin{eqe}\label{eq:SC:5}
\mathcal{A}_i\rhd\mathcal{B}_j.
\end{eqe}%
Thereafter, the basis vector fields of $\mathcal{B}_j$ must be simplified using the normalizer group $\operatorname{Nor}(\mathcal{A}_i, \exp\mathcal{A})$ of $\mathcal{A}_i$. This ensures that the component $\mathcal{A}_i$ is not modified, since its expression has already been simplified in the previous step. The subalgebras of a semi-direct sum of the second type are called nonsplitting subalgebras. They are the subalgebras that cannot be conjugate to splitting ones. They are built from the splitting subalgebras by adding a general component of the complementary space $\mathcal{C}_j$ of $\mathcal{B}_j$ in $\mathcal{B}$ to each basis vector field of the subalgebra $\mathcal{A}_i$ into a splitting subalgebra $\mathcal{F}_{ij}=\mathcal{A}_i\rhd \mathcal{B}_j$. Next, its expression has to be simplified as much as possible through the action of the normalizer group $\operatorname{Nor}(\mathcal{F}_{ij},\exp\mathcal{F})$. If all components in $\mathcal{C}_j$ can be canceled out, then the subalgebra is conjugate to a splitting one. Otherwise the subalgebra is nonsplitting. The key point in this analysis is to compute the action by conjugation on subalgebras made by the group elements. The following subsection will illustrate this notion.
\subsection{Conjugation of subalgebras under the action of a group element}
 Denote $\mathcal{L}$ the Lie algebra associated with the Lie group $G$ of point transformations. Suppose that $\gamma\in\mathcal{L}$ is the vector field generating the transformation $g$ through exponentiation. The vector field $\gamma$ is called the infinitesimal generator of $g$ and it will be denoted $g=\exp \gamma=\sum_{n=0}^\infty \gamma^n/n!$. The action by conjugation of $g$ on a general element $\lambda\in \mathcal{L}$ of the Lie algebra is
\begin{eqe}\label{eq:SC:6}
g\lambda g^{-1}=e^{\gamma}\lambda e^{-\gamma}.
\end{eqe}%
 Through an application of the Baker-Campbell-Hausdorff (BCH) formula, the right hand side of the equation (\ref{eq:SC:6}) can be written as
\begin{eqe}\label{eq:SC:7}
g\lambda g^{-1}=\sum_{n=0}^\infty\frac{[^{(n)}\gamma,\lambda]}{n!},
\end{eqe}%
where the notation
\begin{gaeq}\label{eq:SC:8}
[^{(0)}\gamma,\lambda]=1,\quad [^{(1)}\gamma,\lambda]=[\gamma,\lambda]=\gamma\lambda-\lambda\gamma,\\
[^{(n+1)}\gamma,\lambda]=[\gamma,[^{(n)}\gamma,\lambda]],
\end{gaeq}%
is used. The formula (\ref{eq:SC:7}) allows one to compute the action by conjugation of the subgroup $\exp\mathcal{A}$ on elements of the Abelian ideal subalgebra $\mathcal{B}$. For example, consider the action of the one-parameter rotation group $\exp(\beta L)$ on an element $X_{f(t)}\in\mathcal{B}$, where $\beta\in \RR$ is the group parameter. By virtue of the commutation relations (\ref{eq:10}) it is found that
\begin{eqe}\label{eq:SC:9}
[^{(n)}L,X_{f(t)}]=\left\{\begin{aligned}
&(-1)^{\frac{n-1}{2}}Y_{f(t)}, &&n\quad\text{odd},\\
&(-1)^{\frac{n}{2}}X_{f(t)}, &&n\quad\text{even},
\end{aligned}\right.
\end{eqe}%
and
\begin{eqe}\label{eq:SC:10}
[^{(n)}L,Y_{g(t)}]=\left\{\begin{aligned}
&(-1)^{\frac{n+1}{2}}X_{g(t)}, &&n\quad\text{odd},\\
&(-1)^{\frac{n}{2}}Y_{g(t)}, &&n\quad\text{even}.
\end{aligned}\right.
\end{eqe}%
Hence, applying the relation (\ref{eq:SC:7}) on an element $X_{f(t)}\in \mathcal{B}$ gives
\begin{aleq}
\exp(\beta L)X_{f(t)}\exp(-\beta L)=&\sum_{n=0}^\infty\frac{[^{(n)}\beta L, X_{f(t)}]}{n!},\\
&=\sum_{n=0}^{\infty}\frac{\beta^{2n}(-1)^n}{2n!}X_{f(t)}+\sum_{n=0}^\infty \frac{\beta^{(2n+1)}(-1)^n}{(2n+1)!}Y_{f(t)}.
\end{aleq}%
Since the two series involved in the previous equation correspond to the Taylor series of the trigonometric functions $\cos\beta$ and $\sin\beta$, it results that
\begin{eqe}\label{eq:SC:11}
\exp(\beta L)X_{f(t)}\exp(-\beta L)=\cos(\beta) X_{f(t)}+\sin(\beta) Y_{f(t)}.
\end{eqe}%
Similarly, the action of $\exp(\beta L)$ on $Y_{g(t)}$ is found to be
\begin{eqe}\label{eq:SC:12}
\exp(\beta L)Y_{f(t)}\exp(-\beta L)=\cos(\beta) Y_{g(t)}-\sin(\beta) X_{g(t)}.
\end{eqe}%
For another example, consider the action by conjugation of the one-parameter group $\exp(\alpha D)$ on elements of $\mathcal{B}$, where $\alpha\in\RR$ is the group parameter. First, consider the action on an element $X_{f(t)}\in\mathcal{B}$ for the monomial function $f(t)=t^m$. Hence, it is taken into account that the action occurs over an element of $\mathcal{B}$ of the form
\begin{eqe}\label{eq:SC:9b}
X_{t^m}=t^m\del_x+mt^{m-1}\del_u+m(m-1)\rho xt^{m-2}\del_\sigma, \quad m\geq 2,
\end{eqe}%
or of the form
\begin{eqe}\label{eq:SC:10b}
X_{t^0}=P_1=\del_x,\quad \text{and}\quad X_t=K_1=t\del_x+\del_u,
\end{eqe}%
when $m$ is respectively equal to 0 and 1. The recursive Lie brackets, as defined in equations (\ref{eq:SC:8}), for the vector field $D$ with the vector fields $X_{t^m}$ are
\begin{eqe}\label{eq:SC:11b}
[^{(n)}D, X_{t^m}]=(m-1)^nX_{t^m},\quad m\geq 0.
\end{eqe}%
By virtue of equation (\ref{eq:SC:7}) and using the relation (\ref{eq:SC:11b}), it is found that the action by conjugation of an element $g=\exp(\alpha D)$ on an element $g=\exp(\alpha D)$ on an element $X_{t^m}\in \mathcal{B}$ is
\begin{aleq}
e^{\alpha D}X_{t^m}e^{-\alpha D}=&\sum_{n=0}^\infty\frac{\alpha^n}{n!}(m-1)^nX_{t^m},\\
=&e^{\alpha(m-1)}X_{t^m}.
\end{aleq}%
Now, suppose that $f(t)$ is an analytical function at a time $t_i$ of interest. Thus the vector field $X_{f(t)}$ can be written in the form
\begin{eqe}\label{eq:SC:13}
X_{f(t)}=\sum_{m=0}^\infty \frac{1}{m!}\pa{f_m t^m\del_x+f'_mt^m\del_u+\rho xf''_m t^m\del_\sigma},
\end{eqe}%
where
\begin{eqe}\label{eq:SC:14}
f_m=f^{(m)}(t_i),\quad f'_m=f^{(m+1)}(t_i),\quad f''_m=f^{(m+2)}(t_i),\quad t_i\in \RR,
\end{eqe}%
by replacing the function $f(t)$ and its derivatives by their respective Taylor series at $t_i$. Since, using the definition (\ref{eq:SC:14}), the equalities $f_{m+1}=f'_m$ and $f_{m+2}=f''_m$ hold, and the vector field $X_{f(t)}$ takes the more concise form
\begin{eqe}\label{eq:SC:15}
X_{f(t)}=\sum_{m=0}^\infty \frac{f_m}{m!}X_{t^m},
\end{eqe}%
where $X_{t^m}$ is as defined in equation (\ref{eq:SC:10b}). Therefore, the result (\ref{eq:SC:13}) can be used to evaluate the action of $\exp(\alpha D)$ on $X_{f(t)}\in\mathcal{B}$ as follows
\begin{aleq}
e^{\alpha D}X_{f(t)}e^{-\alpha D}=&\sum_{m=0}^\infty \frac{f_m}{m!} e^{\alpha D} X_{t^m}e^{-\alpha D}
=\sum_{m=0}^\infty\frac{f_m}{m!}e^{\alpha(m-1)}X_{t^m},\\
=&\sum_{m=0}^\infty e^{-\alpha}\frac{f_m}{m!}(e^{\alpha}t)^m\del_x+\sum_{m=1}^\infty \frac{f_m}{(m-1)!}(e^\alpha t)^{m-1}\del_u\\
&+\sum_{m=2}^\infty\frac{f_m}{(m-2)!}e^{\alpha}((e^\alpha)t)^{m-2}\rho x\del_\sigma,\\
&=\sum_{m=0}^\infty e^{-\alpha}\frac{f_m}{m!}(e^{\alpha}t)^m\del_x+\sum_{m=0}^\infty \frac{f'_m}{m!}(e^\alpha t)^{m}\del_u\\
&+\sum_{m=0}^\infty\frac{f''_m}{m!}e^{\alpha}((e^\alpha)t)^{m}\rho x\del_\sigma,\\
=&e^{-\alpha}f(e^\alpha t)\del_x+f'(e^\alpha t)\del_u+\rho x e^\alpha f''(e^\alpha t)\del_\sigma.
\end{aleq}%
Since $f'(e^\alpha t)=e^{-\alpha}\dot{f}(e^\alpha t)$ and $f''(e^\alpha t)=e^{-2\alpha}\ddot{f}(e^\alpha t)$, we have
\begin{eqe}\label{eq:SC:16}
e^{\alpha D}X_{f(t)}e^{-\alpha D}=e^{-\alpha} X_{f(e^\alpha t)}.
\end{eqe}%
Proceeding by analogy, we obtain that
\begin{eqe}\label{eq:SC:17}
e^{\alpha D}Y_{g(t)}e^{-\alpha D}=e^{-\alpha} Y_{g(e^\alpha t)}.
\end{eqe}%
A similar analysis shows that the action of the group element $\exp(t_0 P_0)$, $t_0\in \RR$, on the elements $X_{f(t)}$ and $Y_{g(t)}$ are, respectively
\begin{eqe}\label{eq:SC:18}
e^{toP_0}X_{f(t)}e^{-t_0P_0}=X_{f(t+t_0)},
\end{eqe}%
and
\begin{eqe}\label{eq:SC:18}
e^{toP_0}Y_{g(t)}e^{-t_0P_0}=Y_{g(t+t_0)}.
\end{eqe}%
It is well known that any Lie group element can be written as successive applications of the exponentiation of the basis elements of the associated Lie algebra (see \textit{e.g.} \cite{Olver:Application_of_Lie}). Consequently, the action by conjugation of a generic element of the group $\exp\mathcal{A}$ on an element $X_{f(t)}\in\mathcal{B}$ takes the form
\begin{eqe}\label{eq:SC:20}
e^{-\alpha}\pa{\cos\beta X_{f(e^\alpha t+t_0)}+\sin\beta Y_{g(e^\alpha t+t_0)}},
\end{eqe}%
where $\alpha$, $\beta$ and $t_0$ are real parameters. Equivalently, the action on an element $Y_{g(t)}$ takes the form
\begin{eqe}\label{eq:SC:21}
e^{-\alpha}\pa{\cos\beta Y_{g(e^\alpha t+t_0)}-\sin\beta Y_{g(e^\alpha t+t_0)}}.
\end{eqe}%
Since $\mathcal{B}$ is Abelian, the elements of $\exp\mathcal{B}$ act trivially on the components in $\mathcal{B}$ of an element of $\mathcal{A}\rhd\mathcal{B}$ and since $\mathcal{B}$ is an ideal, all the commutator relations $[^{(n)}\gamma, \lambda]$ in (\ref{eq:SC:7}) vanish for $n>1$. Hence, the series (\ref{eq:SC:7}) contains only two non-vanishing terms for $n=0$ and $n=1$. The term corresponding to $n=1$ is called the cobord. The action of $\exp\mathcal{B}$ on the basis components of $\mathcal{A}$ are
\begin{gaeq}\label{eq:SC:22}
e^{X_{f(t)}}Le^{-X_{f(t)}}=L-Y_{f(t)},\qquad e^{Y_{f(t)}}Le^{-Y_{f(t)}}=L-Y_{f(t)}\\
\begin{aligned}
e^{X_{f(t)}}De^{-X_{f(t)}}=&D+X_{\tilde{f}(t)},\quad &&\tilde{f}(t)=f(t)-t\dot{f}(t),\\
e^{Y_{g(t)}}De^{-Y_{g(t)}}=&D+Y_{\tilde{g}(t)},\quad &&\tilde{g}(t)=g(t)-t\dot{g}(t),
\end{aligned}\\
e^{X_{f(t)}}P_0e^{-X_{f(t)}}=P_0-X_{\dot{f}(t)},\qquad e^{Y_{g(t)}}P_0e^{-Y_{g(t)}}=P_0-Y_{\dot{g}(t)}.
\end{gaeq}%
One should note that $\tilde{f}(t)=f(t)-t\dot{f}(t)$ and $\tilde{g}(t)=g(t)-t\dot{g}(t)$ are ODEs that always possess a solution for $f(t)$ and $g(t)$ for any given analytical functions $\tilde{f}(t)$ and $\tilde{g}(t)$. This means that $\tilde{f}(t)$ and $\tilde{g}(t)$ can be chosen as desired in the cobord of $D$ in equation (\ref{eq:SC:22}). Consequently, this allows us to simplify a lot the representative subalgebras of the conjugacy classes. In fact, as one can see from the results of the next subsection, there are sufficiently many possibilities of conjugation with the cobords (\ref{eq:SC:22}) to ensure that there exist no nonsplitting subalgebras of $\mathcal{F}=\mathcal{A}\rhd\mathcal{B}$ that are not conjugate to a splitting one.
\subsection{Results of the classification of $\mathcal{F}=\mathcal{A}\rhd\mathcal{B}$}
First, consider the one-dimensional splitting subalgebras. They may be either of the form $\mathcal{A}_{1,i}\rhd\langle0\rangle\cong\mathcal{A}_{1,i}$, with $\mathcal{A}_{1,i}$ a one-dimensional subalgebra of $\mathcal{A}$, or either of the form $\langle 0\rangle\rhd \mathcal{B}_{1,i}$, with $\mathcal{B}_{1,i}$ a one-dimensional subalgebra of $\mathcal{B}$. The symbol $\langle 0\rangle$ denotes the subalgebra consisting of the identity only. In the first case, the representative subalgebras of conjugacy classes are listed in table \ref{tab:1}. For the second case the subalgebras are of the form
\begin{eqe}\label{eq:SC:23}
\langle X_{f(t)}+Y_{g(t)}\rangle,
\end{eqe}%
where $f(t)$ and $g(t)$ are assumed to be analytical functions in an interval around a time $t_i$ of interest and the vector fields $X_{f(t)}$ and $Y_{g(t)}$ are defined in (\ref{eq:SC:15}). It is supposed that the Wronskian
$$\left|\begin{array}{cc}
          f(t) & g(t) \\
          \dot{f}(t) & \dot{g}(t)
        \end{array}
\right|$$%
does not vanish, otherwise a subalgebra of the form (\ref{eq:SC:23}) can be brought to the simpler form
\begin{eqe}\label{eq:SC:24}
\langle X_{f(t)}\rangle
\end{eqe}%
using conjugation under the action of a group element $\exp(\beta L)$, $\beta\in \RR$. The form (\ref{eq:SC:24}) is just a special case of the form (\ref{eq:SC:23}) when $f(t)=h(t)$ and $g(t)=0$. Now, assume that the function $f(t)$ has a zero at $t_z$ in the interval around $t_i$ where the Taylor series of $f(t)$ converges. Using the conjugation under the action of the one-parameter group $\exp(t_0P_0)$, $t_0\in \RR$, the subalgebra (\ref{eq:SC:23}) is in the same conjugacy class as the subalgebra $\langle X_{\tilde{h}(t)}+Y_{\tilde{g}(t)}\rangle$, where $\tilde{h}(t)$ is an analytical function with a zero at $t=0,\textit{i.e.}$ $\tilde{h}(t)=f(t+t_z)$. Moreover, since $\langle X_{c\tilde{h}(t)}+Y_{c\tilde{g}(t)}\rangle\cong \langle c(X_{\tilde{h}(t)}+Y_{c\tilde{g}(t)})\rangle\cong \langle X_{\tilde{h}(t)}+Y_{\tilde{g}(t)}\rangle$, $c\in \RR$, we can divide the vector field spanning the one-dimensional subalgebra (\ref{eq:SC:23}) by the value of the first non-vanishing derivative at $t=0$. This implies that the subalgebra (\ref{eq:SC:23}) is conjugate to subalgebra
\begin{eqe}\label{eq:SC:26}
\langle X_{h(t)}+Y_{g(t)}\rangle,
\end{eqe}%
where $h(t)=\tilde{h}(t)/h^{(m_1)}(0)$ and $\hat{g}(t)=\tilde{g}(t)/h^{(m_1)}(0)$, with $1\leq m_1\in \NN$, being the order of the first non-vanishing derivatives at $t=0$. Consequently, the Taylor series of $h(t)$ takes the form
\begin{eqe}\label{eq:sc:27}
h(t)=t^{m_1}+\sum_{n=m_1+1}^\infty \frac{h^{(n)}(0)}{n!}t^n,\quad 1\leq m_1.
\end{eqe}%
Next, acting by conjugation with a group element $e^{\alpha D}$ on (\ref{eq:SC:26}) and dividing the vector field by $e^{\alpha m_1}$, we see that the subalgebra remains of the form (\ref{eq:SC:26}) but the function $h(t)$ have the Taylor series given by
$$h(t)=t^{m_1}+\frac{e^{\alpha(m_2-m_1)} h^{(m_2)}(0)}{m_2!}t^{m_2}+\sum_{n=m_2+1}^\infty \frac{e^{\alpha (n-m_1)}h^{(n)}(0)}{n!}t^n.$$%
 However, the coefficient of the monomial term in $t$ of degree equal to $m_2$ can be set to $\pm 1$ by the choice $e^{\alpha(m_2-m_1)}=|h^{(m_2)}(0)/m_2!|$. Thus, if a one-dimensional subalgebra of the form (\ref{eq:SC:23}), with $f(t)$ being an analytical function at $t_i$ and having a zero in the interval of convergence, then it is conjugated to a subalgebra which is again of the form (\ref{eq:SC:23}) but where the functions $f(t)$ and $g(t)$ have the form
\begin{aleq}\label{eq:SC:28}
f(t)=&t^{m_1}+\mu\pa{t^{m_2}+t^{m_2+1}\tilde{f}(t)},\quad 1\leq m_1,\quad m_1+|\mu|\neq0\\
g(t)=&\tilde{g}(0)^{-|\mu|}\tilde{g}(t),
\end{aleq}%
where $\mu\in\ac{-1,0,1}$, and the functions $\tilde{f}(t)$ and $\tilde{g}(t)$ are arbitrary analytical functions.  When $\mu=0$, the arbitrariness of $g(t)=\tilde{g}(t)$ is due to the fact that once we have simplified the component $X_{f(t)}$ in the vector field (\ref{eq:SC:23}), the component $X_{g(t)}$ must be simplified with the normalizer group of $X_{f(t)}$ in $\exp\mathcal{A}$, which consist of the identity only. So no further simplifications of $g(t)$ are possible. The only other possible case occurs when $f(t)$ is a constant function $(m_1=\mu=0)$. The normalizer group of $X_{f(t)}$ is then $\exp(\mathcal{A})$ and the subalgebra (\ref{eq:SC:23}) is conjugated to one of the same form but with functions $f(t)$ and $g(t)$ defined by
$$
f(t)=1,\qquad g(t)=t^{m_1}+\mu (t^{m_2}+t^{m_2+1}\tilde{g}(t)),
$$
where $\tilde{g}(t)$ is an arbitrary analytical function in the neighborhood of $t=0$.
\paragraph{}Now, in the case where the function $f(t)$ in (\ref{eq:SC:23}) does not have a zero in the interval of convergence, it suffices to remark that the function $\tilde{f}(t)=(t-t_i)f(t)$ will have one. This implies that the Taylor series of $f(t)$ at $t=t_i$ is
\begin{eqe}\label{eq:SC:29}
f(t)=\frac{\tilde{f}^{(m_1)}(t_i)}{m_1!}(t-t_i)^{m_1-1}+\frac{\tilde{f}^{(m_2)}(t_i)}{m_2!}(t-t_i)^{m_2-1}+\sum_{n=m_2+1}^{\infty}\frac{\tilde{f}(t_i)}{n!}(t-t_i)^{n-1},
\end{eqe}%
$1\leq m_1\leq m_2$. Now, acting by conjugation with the group element $\exp(t_iP_0)$, the subalgebra (\ref{eq:SC:23}) is brought to the subalgebra $\langle X_{\tilde{h}(t)}+Y_{\tilde{g}(t)}\rangle$ where $\tilde{h}(t)=\frac{f(t+t_i)}{\tilde{f}^{(m_1)}(t_i)}m_1!$. So, the Taylor series of $\tilde{h}(t)$ is
$$\tilde{h}(t)=t^{m_1-1}+\frac{\tilde{h}^{(m_2)}(0)}{m_2!}t^{m_2-1}+\sum_{n=m_2+1}^{\infty}\frac{\tilde{h}(0)}{n!}t^{n},\quad 1\leq m_1\leq m_2.$$%
Thereafter, acting with the appropriate element of $\exp\langle D\rangle$ it is found that the subalgebra (\ref{eq:SC:23}) is finally conjugate to
\begin{eqe}\label{eq:SC:30}
\langle X_{h(t)}+Y_{g(t)}\rangle,
\end{eqe}%
where
\begin{aleq}\label{eq:SC:31}
h(t)=&t^{m_1}+\mu\pa{t^{m_2}+t^{m_2+1}\hat{h}(t)},\quad 0\leq m_1\leq m_2,\quad \mu\in\ac{-1,0,1},\\
g(t)=&\hat{g}(0)^{-|\mu|}\hat{g}(t), \quad m_1+|\mu|\neq 0,
\end{aleq}%
or
\begin{eqe}\label{eq:SC:31b}
h(t)=1,\quad
g(t)=t^{m_3}+\mu\pa{t^{m_4}+t^{m_4+1}\hat{g}(t)},\quad 1\leq m_3\leq m_4,\\
\end{eqe}%
and $\hat{g}(t)$ and $\hat{h}(t)$ are analytical functions at $t=0$. In summary, the representatives of the one-dimensional splitting subalgebras classes, not listed in table \ref{tab:1}, are those of the form (\ref{eq:SC:30}) with $h(t)$ and $\hat{g}(t)$ given by (\ref{eq:SC:31}) or (\ref{eq:SC:31b}). Turning to the case of the nonsplitting subalgebras, it is easily verified using the cobords (\ref{eq:SC:22}) that all such subalgebras are conjugate to splitting ones.
\paragraph{}Concerning the two-dimensional subalgebras there are three possibilities. They are of the type $\langle 0\rangle \rhd \mathcal{B}_{2,j}$, $\mathcal{A}_{1,i}\rhd \mathcal{B}_{1,j}$ or $\mathcal{A}_{2,i}\rhd \langle 0\rangle$, where the first index in the notation of the subalgebra denotes the dimension. The latter type has already been classified in table \ref{tab:1}. For the subalgebras of the type $\langle 0\rangle \rhd \mathcal{B}_{2,j}$, the conjugacy classes are represented by the following subalgebra
\begin{eqe}\label{eq:SC:32}
\langle X_{f_1(t)}+Y_{g_1(t)},X_{f_2(t)}+Y_{g_2(t)}\rangle,
\end{eqe}%
where the analytical functions $f_1(t)$, $g_1(t)$, $f_2(t)$ and $g_2(t)$ have the form
\begin{gaeq}\label{eq:SC:33}
f_1(t)=t^{m_1}+\mu\pa{t^{m_2}+t^{m_2+1}\hat{f}_1(t)},\quad 0\leq m_1<m_2,\\
g_1(t)=\hat{g}(0)^{-|\mu|}\hat{g}_1(t),\quad f_2(t)=\hat{f}_2(t),\quad g_2(t)=\hat{g}_2(t),
\end{gaeq}%
or
\begin{gaeq}\label{eq:SC:33b}
f_1(t)=1,\quad g_1(t)=t^{m_3}+\mu\pa{t^{m_4}+t^{m_4+1}\hat{g}_1(t)},\quad 1\leq m_3<m_4,\\
\quad f_2(t)=\hat{f}_2(t),\quad g_2(t)=\hat{g}_2(t),
\end{gaeq}%
or
\begin{gaeq}\label{eq:SC:33c}
f_1(t)=1,\quad g_1(t)=0\\
\quad f_2(t)=t^{m_1}+\mu\pa{t^{m_2}+t^{m_2+1}\hat{f}_2(t)},\quad 0\leq m_1<m_2,\quad g_2(t)=\hat{g}_2(0)^{-|\mu|}\hat{g}_2(t),
\end{gaeq}%
or
\begin{gaeq}\label{eq:SC:33d}
f_1(t)=1,\quad g_1(t)=0,\\
\quad f_2(t)=1,\quad g_2(t)=t^{m_3}+\mu\pa{t^{m_4}+t^{m_4+1}\hat{g}_2(t)},\quad 1\leq m_3<m_4,
\end{gaeq}%
where $\hat{f}_1(t)$, $\hat{g}_1(t)$, $\hat{f}_2(t)$ and $\hat{g}_2(t)$ are arbitrary analytical functions of $t$. Concerning subalgebras of the type $\mathcal{A}_{1,i}\rhd \mathcal{B}_{1,j}$, one has to find, for each one-dimensional subalgebra listed in Table \ref{tab:1}, a one-dimensional subalgebra of $\mathcal{B}$ that forms an ideal. Thereafter, the parameters have to be simplified as much as possible using conjugacy under $\operatorname{Nor}(\mathcal{A}_i,\exp\mathcal{A})$. The resulting two-dimensional splitting subalgebras are listed in Table \ref{tab:2}.
\begin{table}
\begin{center}
\begin{tabular}{ll}
  \hline
  representative subalgebra &  normalizer subgroup\\
  $\langle D,X_{t^a}\rangle$ & $\exp\langle D\rangle$ \\
  $\langle P_0,X_{e^{\pm t}}\rangle$ & $\exp\langle P_0\rangle$ \\
  $\langle L+aD,X_{t^b}\cos(a^{-1}\ln t)-Y_{t^b}\sin(a^{-1}\ln t)\rangle,\ a\neq0$ & $\exp(\pi L)\times \operatorname{Nor}(L+aD,\exp\mathcal{B})$ \\
  $\langle L+P_0, X_{e^{at}}\sin t+Y_{e^{at}}\cos t\rangle$ & $\exp(\pi L)\times \operatorname{Nor}(L+P_0,\exp\mathcal{B})$ \\
  \hline
\end{tabular}
\end{center}
\caption{List of two-dimensional representative subalgebras of the type $\mathcal{A}_{1,i}\rhd \mathcal{B}_{1,j}$. Here $a$ and $b$ are real parameters.}\label{tab:2}
\end{table}%
It is easily checked that all two-dimensional nonsplitting subalgebras of $\mathcal{F}$ is conjugate to one of the list in table \ref{tab:2}. Hence, the classification of classes of one- and two-dimensional subalgebras of $\mathcal{F}$ are complete.
\paragraph{}Proceeding in a similar way, this classification can be extended to a classification $\mathcal{L}=\mathcal{F}\rhd \mathcal{S}$. Here, only the results are given. To the previously found representatives of conjugacy classes, the following ones must be added
\begin{gaeq}\label{eq:classResult}
\langle S_{h(t)} \rangle,\quad \langle S_1,S_{h(t)} \rangle,\quad \langle S_{t^{m_1}}, S_{f(t)} \rangle,\quad \langle S_{h(t)},S_{g(t)} \rangle,\\
\langle L,S_{h(t)}\rangle, \quad\langle D,S_1\rangle,\quad \langle D,S_{t^a}\rangle,\quad \langle P_0,S_1\rangle,\quad \langle P_0,S_{e^{\pm t}}\rangle,\\
\langle L+aD, S_1\rangle,\quad \langle L+aD,S_{t^a}\rangle,\quad \langle L+aD,X_{t^b\cos(a^{-1}\ln t)}-Y_{t^b\sin(a^{-1}\ln t)+c S_{t^b}}\rangle,\\
\langle L+P_0, S_1\rangle,\quad \langle L+P_0,S_{e^{\pm t}}\rangle,\quad \langle L+aD,X_{\exp(at)\sin t}+Y_{\exp(at)\cos t+c S_{\exp(at)}}\rangle.
\end{gaeq}%
where $a\neq 0$, $b$ and $c$ are real parameters,
$$f(t)=t^{n_1}+\epsilon t^{n_2}+t^{n_2+1}\hat{g}(t),\quad h(t)=t^{m_1}+\mu\pa{t^{m_2}+t^{m_2+1}\hat{h}(t)},\quad 0\leq m_1<m_2,$$%
and $g$, $\hat{g}$ and $\hat{h}$ are arbitrary analytical functions such that $h\dot{g}-\dot{h}g\neq 0$.
\section{Symmetry reductions}\label{sec:4}
In this section we illustrate the use of the results obtained in the previous section by several examples of invariant solutions for different types of forces considered in Section \ref{sec:2}. The symmetry reduction method (see \textit{e.g.} \cite{Olver:Application_of_Lie}) is used to obtain solutions of the system (\ref{eq:1}).
\subsection{Monogenic forces}
\begin{figure}[h]
\begin{center}
\includegraphics[width=2in]{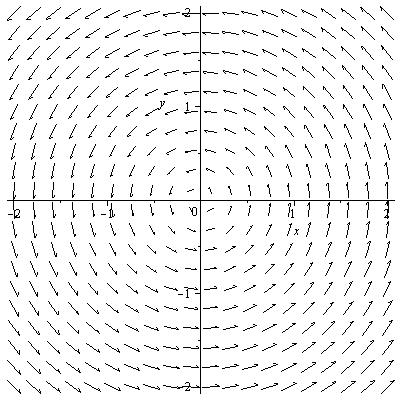}
\includegraphics[width=2in]{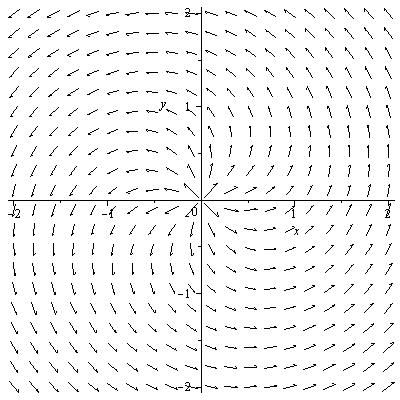}
\includegraphics[width=2in]{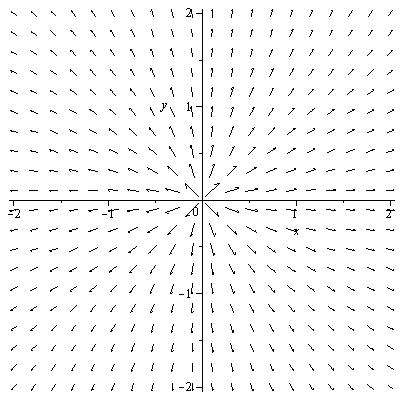}
\end{center}
\caption{Evolution of the vectors field of the solution (\ref{eq:R:17}).}\label{fig:1}
\end{figure}

Let us consider the system (\ref{eq:1}) when the force involved is monogenic, \ie of the form (\ref{eq:p:4}). In this case the system (\ref{eq:1}) admits the largest symmetry group and the classification of its subalgebras into conjugacy classes has been performed in the previous section for subalgebras of dimension 1 and 2. As a first example, we obtain particular solutions which are invariant under the action of the subgroup spanned by the two-dimensional subalgebra
\begin{eqe}\label{eq:R:1}
\mathcal{L}_{2,1}=\ac{D,L},
\end{eqe}%
when the parameters are $a=b=0$ and where the generators $D$ and $L$ are defined in equation (\ref{eq:9}). This subalgebras admits the following functionally independent invariants
\begin{gaeq}\label{eq:R:2}
\xi=\frac{x^2+y^2}{t^2},\quad T_1=\theta-\arctan\pa{\frac{y}{x}},\quad T_2=\theta-\arctan\pa{\frac{v}{u}},\\
R=u^2+v^2,\quad S=\sigma+\rho V(t,x,y),
\end{gaeq}%
where $V(t,x,y)$ is a function which defines the monogenic force. The last four relations in (\ref{eq:R:2}) can be inverted to obtain $u,v,\theta,\sigma$ in terms of $t,x,y$ and of the invariants $T_1,T_2,R,S$. Therefore, assuming that the invariants $T_1$, $T_2$, $R$ and $S$ are functions of the invariant $\xi$, called the symmetry variable, we make the hypothesis that the invariant solution is of the form
\begin{aleq}\label{eq:R:3}
u=&R(\xi)^{1/2}\cos\pa{T_1(\xi)-T_2(\xi)+\arctan(y/x)},\qquad &&\theta=T_1(\xi)+\arctan(y/x),\\
v=&R(\xi)^{1/2}\sin\pa{T_1(\xi)-T_2(\xi)+\arctan(y/x)}, &&\sigma=S(\xi)-\rho V(t,x,y).\\
\end{aleq}%
Introducing the Ansatzes (\ref{eq:R:3}) into the system (\ref{eq:1}) for the monogenic force (\ref{eq:p:4}), we obtain the following reduced system
\begin{aleq}\label{eq:mred1}
S'(\xi)=&\frac{\rho}{2}\pa{\frac{1}{2}\pa{\cos\pa{2T_1(\xi)-2T_2(\xi)}+1}-(\xi/R(\xi))^{1/2}\cos\pa{T_1(\xi)-T_2(\xi)}}R'(\xi)\\
&+\rho\xi R(\xi)\bigg(\cos(2T_1(\xi))-2^{-1}R(\xi)\rho\sin\pa{2T_1(\xi)-2T_2(\xi)}\\
&+\rho\pa{\xi R(\xi)}^{1/2}\sin\pa{T_1(\xi)-T_2(\xi)}\bigg)T_1'(\xi)+\frac{1}{2\xi}\sin\pa{2T_1(\xi)}-\frac{\rho R(\xi)}{4\xi}\\
&+\rho\Big(\frac{R(\xi)}{2}\sin\pa{2T_1(\xi)-2T_2(\xi)}-\pa{\xi R(\xi)}^{1/2}\sin\pa{T_1(\xi)-T_2(\xi)}\Big)T_2'(\xi)\\
&+\frac{\rho R(\xi)}{4\xi}\cos\pa{2T_1(\xi)-2T_2(\xi)},
\end{aleq}%
\begin{aleq}\label{eq:mred2}
&\rho\xi\pa{\frac{1}{2}\sin\pa{2T_1(\xi)-2T_2(\xi)}-\pa{\xi /R(\xi)}^{1/2}\sin\pa{T_1(\xi)-T_2(\xi)}}R'(\xi)\\
&+\rho\xi\Big(R(\xi) \cos\pa{2T_1(\xi)-2T_2(\xi)}-2\pa{\xi R(\xi)}^{1/2}\cos\pa{T_1(\xi)-T_2(\xi)}\\
&+2\sin(T_1(\xi))+R(\xi)\Big)T_1'(\xi)-\cos\pa{2T_1(\xi)}+\rho\xi R(\xi)\Big(-\cos\pa{2T_1(\xi)-2T_2(\xi)}\\
&+2\pa{\xi/R(\xi)}^{1/2}\cos\pa{T_1(\xi)-T_2(\xi)}-1\Big)T_2'(\xi)+\frac{\rho}{2}R(\xi)\sin\pa{2T_1(\xi)-2T_2(\xi)}=0,
\end{aleq}%
\begin{aleq}\label{eq:mred3}
&\frac{1}{2}\pa{\xi+\cos\pa{2T_1(\xi)+2T_2(\xi)}}R'(\xi)+\xi R(\xi)\sin\pa{2T_1(\xi)+2T_2(\xi)}T_1'(\xi)\\
&-\xi R(\xi)\sin\pa{2T_1(\xi)+2T_2(\xi)}T_2'(\xi)-\frac{R(\xi)}{2}\pa{1+\cos\pa{2T_1(\xi)+2T_2(\xi)}}=0,
\end{aleq}%
\begin{aleq}\label{eq:mred4}
&\xi\cos\pa{T_1(\xi)-T_2(\xi)}R'(\xi)-2\xi R(\xi)\sin\pa{T_1(\xi)-T_2(\xi)}T_1'(\xi)\\
&+2\xi R(\xi)\sin\pa{T_1(\xi)-T_2(\xi)}T_2'(\xi)+R(\xi)\cos\pa{T_1(\xi)-T_2(\xi)}=0.
\end{aleq}%
It should be noted that the first equations (\ref{eq:mred1}) and (\ref{eq:mred2}) are found by taking the combination $x\cdot (\ref{eq:1}.a)+y\cdot(\ref{eq:1}.b)$ and $y\cdot(\ref{eq:1}.a)-x\cdot(\ref{eq:1}.b)$, and then substituting the Ansatzes (\ref{eq:R:3}) into those combinations. The reduced equation (\ref{eq:mred4}) admits the first integral
\begin{eqe}\label{eq:R:4}
(1/2)\xi R(\xi)\pa{1+\cos\pa{(2T_1(\xi))-2T_2(\xi)}}=a_1,
\end{eqe}%
where $a_1$ is a real integration constant. So, solving (\ref{eq:R:4}) for $T_1(\xi)$ we find
\begin{aleq}\label{eq:R:5}
T_1(\xi)=&T_2(\xi)+\frac{\epsilon}{2}\arccos\pa{\frac{2a_1}{\xi R(\xi)}-1},\\
=&T_2(\xi)+\arccos\pa{\frac{\epsilon a_1^{1/2}}{(\xi R(\xi))^{1/2}}},\quad \epsilon=\pm 1.
\end{aleq}%
Substituting (\ref{eq:R:5}) into the reduced equation (\ref{eq:mred2}), we obtain an equation relating the quantities $R$ and $T_2$ and their first derivatives. From this equation we obtain the first integral
\begin{aleq}\label{eq:R:6}
&-\frac{1}{2}a_1^{-1/2}\xi\pa{1-\frac{2a_1}{\xi R(\xi)}}\cos(2T_2(\xi))-\frac{(\xi R(\xi)-a_1)^{1/2}}{R(\xi)}\sin\pa{2T_2(\xi)}\\
&+\rho\pa{a_1^{1/2}\xi-a_1}a_1^{-1}(\xi R(\xi)-a_1)^{1/2}-\frac{3}{2}a_1^{-1/2}\rho\int\pa{\xi R(\xi)-a_1}^{1/2} d\xi=a_2,
\end{aleq}%
where $a_2$ is a real integration constant and we have chosen $\epsilon=-1$. Replacing (\ref{eq:R:5}) into (\ref{eq:mred3}) leads to the following ODE for $R(\xi)$
\begin{eqe}\label{eq:R:7}
-2 a_1^{1/2}\pa{\xi R(\xi)-a_1}^{1/2}R'(\xi)\cos\pa{2 T_2(\xi)}+\pa{\pa{2a_1-\xi R(\xi)}R'(\xi)+R(\xi)^2}\sin\pa{T_2(\xi)}=0.
\end{eqe}%
The function $S(\xi)$ is obtained by integrating the equation (\ref{eq:mred1}) and can be expressed in terms of the functions $R$ and $T_2$ in the form
\begin{aleq}\label{eq:defS}
S(\xi)=&-\frac{1}{2}\frac{\pa{\xi R(\xi)-2a_1}\sin(2T_2(\xi))}{\xi R(\xi)}+\frac{a_1^{1/2}\pa{\xi R(\xi)-a_1}^{1/2}\cos(2T_2(\xi))}{\xi R(\xi)}\\
&-\int\frac{1}{2\xi}\pa{\frac{R(\xi)\sin(2T_2(\xi))}{\xi R'(\xi)}+\rho\pa{\xi R(\xi)-a_1^{1/2}}}d\xi+a_3,
\end{aleq}%
where $a_3$ is a real integration constant. Therefore, if we obtain the functions $R$ and $T_2$ which satisfy equations (\ref{eq:R:6}) and (\ref{eq:R:7}) then those functions, together with $T_1$ and $S$ defined respectively by (\ref{eq:R:5}) and (\ref{eq:defS}), provide a solution of the system (\ref{eq:1}) when they are replaced into the formulas (\ref{eq:R:3}) (when the force is of the form (\ref{eq:p:4})). Equations (\ref{eq:R:6}) and (\ref{eq:R:7}) are difficult to solve in general, but certain particular solutions can be obtained. For example, equations (\ref{eq:R:6}) and (\ref{eq:R:7}) are satisfied by the particular solution
\begin{eqe}\label{eq:R:8}
R(\xi)=a_1\xi^{-1},\quad T_2(\xi)=\pi/4,\quad a_2=0.
\end{eqe}%
In this case, substituting (\ref{eq:R:8}) into (\ref{eq:R:5}) and (\ref{eq:defS}), we find respectively that
\begin{eqe}\label{eq:R:9}
T_1(\xi)=\pi/4,\quad S(\xi)=\frac{1}{2}\pa{\rho a_1^{1/2}\ln(\xi)+\ln\pa{\xi+\pa{\xi^2-4a_1a_2^2}^{1/2}}+\rho a_1 \xi^{-1}}+a_3.
\end{eqe}%
Finally, replacing (\ref{eq:R:8}) and (\ref{eq:R:9}) into (\ref{eq:R:3}) we find the explicit solution in the form
\begin{gaeq}\label{eq:R:10}
\theta=\pi/4+\arctan(y/x),\quad u=\frac{a_1^{1/2}tx}{x^2+y^2},\quad v=\frac{a_1^{1/2}ty}{x^2+y^2},\\
\sigma=\pa{\rho a_1^{1/2}+1}\ln\pa{\frac{(x^2+y^2)^{1/2}}{t}}+\frac{a_1 \rho}{2}\frac{t^2}{x^2+y^2}-\rho V(t,x,y)+a_3,
\end{gaeq}%
where $V(t,x,y)$ is the function which defines the force (\ref{eq:p:4}). This solution is irrotational since it makes the quantity $u_y-u_x$ vanish. Consequently, this solution is a particular case of the solutions constructed in \cite{GrundlandLamothe:2013}, where all irrotational solutions were obtained for the case of a monogenic force. Another particular solution can be obtained when $a_1=0$ in the first integral (\ref{eq:R:4}). In this case, the functions $T_1$ can be expressed in terms of $T_2$ in the form
\begin{eqe}\label{eq:R:11}
T_1(\xi)=T_2(\xi)-\pi/2.
\end{eqe}%
As a consequence of (\ref{eq:R:11}) and of $a_1=0$, equation (\ref{eq:R:7}) implies that
$$\xi R'(\xi)-R(\xi)=0,$$%
which is satisfied by
\begin{eqe}\label{eq:R:12}
R(\xi)=b_1^2\xi,
\end{eqe}%
where $b_1$ is an integration constant. So replacing (\ref{eq:R:11}) and (\ref{eq:R:12}) into (\ref{eq:R:3}), we see that the solution must be of the form
\begin{eqe}\label{eq:R:13}
\theta=T_1(\xi)+\arctan(y/x),\quad \sigma=S(\xi)-\rho V(t,x,y),\quad u=-\frac{b_1 y}{t},\quad v=\frac{b_1 x}{t}.\\
\end{eqe}%
We introduce this solution into the first two equations of (\ref{eq:1}) in order to obtain the reduced equations
\begin{gaeq}\label{eq:R:14}
S'(\xi)=\cos\pa{2T_1(\xi)}T_1'(\xi)+\frac{1}{2}\frac{\sin\pa{2T_1(\xi)}}{\xi}-\frac{1}{2}\rho b_1^2,\\
2\xi\sin\pa{2T_1(\xi)}T_1'(\xi)-\cos\pa{2T_1(\xi)}-\rho b_1\xi=0,
\end{gaeq}%
for the functions $S$ and $T_1$. The solution of system (\ref{eq:R:14}) is given by
\begin{aleq}\label{eq:R:15}
T_1(\xi)=&\frac{\pi}{2}-\frac{1}{2}\arccos\pa{\frac{\rho b_1\xi^2+2b_2}{\xi}},\\
S(\xi)=&-\frac{1}{2}\int^{\eta=\xi}\frac{(\rho b_1\eta)^2+2(\rho b_1 b_2-1)}{\pa{4\eta^2-\pa{\rho b_1 \eta^2+2 b_2}^2}^{1/2}}d\eta-\frac{\rho b_1^2}{2}\xi+b_3,
\end{aleq}%
where $b_i$, $i=1,2,3$, are real integration constants. The corresponding solution of system (\ref{eq:1}) takes the form
\begin{gaeq}\label{eq:R:16}
u=-\frac{b_1 y}{t},\quad v=\frac{b_1x}{t},\\
\theta=\frac{\pi}{2}-\frac{1}{2}\arccos\pa{\frac{1}{2}\frac{\rho b_1\pa{x^2+y^2}+2b_1 t^4}{t^2 (x^2+y^2)}}+\arctan(y/x),\\
\sigma=-\rho V(t,x,y)-\frac{1}{2}\rho\frac{b_1^2(x^2+y^2)}{t^2}+\int^{\eta=(x^2+y^2)/t^2}\frac{(\rho b_1\eta)^2+2(\rho b_1b_2-1)}{\pa{4\eta^2-\pa{\rho b_1\eta^2+2b_2^2}^2}^{1/2}}+b_3.
\end{gaeq}%
\paragraph{}From the previous solutions ((\ref{eq:R:10}) and (\ref{eq:R:16})), it is possible to find an additional solution. First, we assume that the components of the velocity form a linear combination of the velocity component in solutions (\ref{eq:R:10}) and (\ref{eq:R:16}), \textit{i.e.}
\begin{eqe}\label{eq:R:17}
u=\frac{a_1^{1/2}tx}{x^2+y^2}-\frac{a_2y}{t},\qquad v=\frac{a_1^{1/2}ty}{x^2+y^2}+\frac{a_2x}{t},
\end{eqe}%
where $a_1,a_2\in\RR$. Next, we replace (\ref{eq:R:17}) into equation (\ref{eq:1}.c), which we then solve algebraically in order to find $\theta$ in the form
\begin{eqe}\label{eq:R:18}
\theta=-\frac{1}{2}\arctan\pa{\frac{1}{2}\frac{x^2-y^2}{xy}}.
\end{eqe}%
It is easily verified that $\theta$ given by (\ref{eq:R:18}) and the velocity components $u$ and $v$ given by (\ref{eq:R:17}) satisfy the compatibility condition of the mixed derivatives of $\sigma$ with respect to $x$ and $y$ when the force takes the form:
\begin{eqe}\label{eq:R:19}
F_1=V_x(t,x,y)+\frac{a_2y}{t},\qquad F_2=V_y(t,x,y)-\frac{a_2 x}{t},
\end{eqe}%
where $V(t,x,y)$ is an arbitrary real-valued function. In this case, $\sigma$ is obtained by integrating equations (\ref{eq:1}.a) and (\ref{eq:1}.b) under the form
\begin{aleq}\label{eq:R:20}
\sigma=&-\rho V(t,x,y)+2^{-1}(a_1\rho-1)\ln\pa{x^2+y^2}-2\rho a_1a_2\arctan(y/x)\\
&-2^{-1}\rho a_2^2 t^{-2}(x^2+y^2)+2^{-1}\rho a_1^2t^2(x^2+y^2)^{-1}+s(t),
\end{aleq}%
where $s(t)$ is an arbitrary function of time. As shown in figure 1, the vector fields (\ref{eq:R:17}) of this solution evolve from a concentric form at the initial times to a radial form at sufficiently large times. During the transition between the two configurations, the flow lines spiral away from the origin. The vector fields have been drawn in Figure \ref{fig:1} for the values $a_1=1$ and $a_2=1$ at the times $t=0.1$, $t=1$ and $t=10$ from left to right. It should be noted that when the parameter $a_1$ changes its sign, the field lines converge to the origin (at sufficiently large times) instead of diverging from it. When the parameter $a_2$ changes its sign, the rotational direction of the flow lines is reversed (at sufficiently small times).
\subsection{Solutions in the presence of a frictional force.}
Consider the system (\ref{eq:1}) when the involved force is of form (\ref{eq:fr:1}). This force can be expressed in terms of two arbitrary functions, which depend on the velocity. This allows us to consider certain problems involving friction. For this type of force the symmetry algebra is spanned by the generators (\ref{eq:fr:2}). We focus on solutions which are invariant under the action of the subgroup corresponding to the one-dimensional subalgebra generated by
$$K=\kappa_1\pa{t\del_t+x\del_x+y\del_y}+\kappa_2\pa{y\del_x-x\del_y+v\del_u-u\del_v-\del_\theta},$$%
where the parameters $\kappa_1$ and $\kappa_2$ appear in the force. A set of functionally independent invariants of the generator $K$ is given by
\begin{gaeq}\label{eq:rf:1}
r=\frac{x^2+y^2}{t^2},\quad \xi=\kappa_2\ln(t)+\kappa_1\arctan(y/x),\quad R=u^2+v^2,\\
T_1=\kappa_2\ln(t)+\kappa_1\theta,\quad T_2=\kappa_2\ln(t)+\kappa_1\arctan(v/u),\quad S=\sigma.
\end{gaeq}%
Assuming that the invariants $R,T_1,T_2$ and $S$ are functions of $r$ and $\xi$, we invert the relations (\ref{eq:rf:1}) in order to obtain $u$, $v$, $\theta$ and $\sigma$ in terms of the invariant solution, \ie
\begin{gaeq}\label{eq:rf:2}
u=R(r,\xi)\cos\pa{\frac{T_2(r,\xi)-\kappa_2\ln(t)}{\kappa_1}},\qquad v=R(r,\xi)\sin\pa{\frac{T_2(r,\xi)-\kappa_2\ln(t)}{\kappa_2}},\\
\theta=T_1(r,\xi)-\frac{\kappa_2}{\kappa_1}\ln(t),\qquad \sigma=S(r,\xi).
\end{gaeq}%
Here, we have made the hypothesis that $\kappa_1\neq0$. Replacing the force (\ref{eq:fr:1}) and the Ansatz (\ref{eq:rf:2}) into the system (\ref{eq:1}), we obtain the reduced system
\begin{aleq}\label{eq:dSr}
\frac{\del S}{\del r}=&\rho\pa{-r^{1/2}\cos\pa{\frac{T_2-\xi}{\kappa_1}}+\frac{R}{2}\pa{1+\cos\pa{\frac{2(T_2-\xi)}{\kappa_1}}}}\frac{\del R}{\del r}\\
&+\frac{\rho}{2}\pa{\frac{\kappa_2}{r^{1/2}}\cos\pa{\frac{T_2-\xi}{\kappa_1}}+\frac{\kappa_1 R}{2r}\sin\pa{\frac{2(T_2-\xi)}{\kappa_1}}}\frac{\del R}{\del \xi}+\cos\pa{\frac{2(\kappa_1T_1-\xi)}{\kappa_1}}\frac{\del T_1}{\del r}\\
&+\frac{\kappa_1}{2 r}\sin\pa{\frac{2(\kappa_1T_1-\xi)}{r}}\frac{\del T_1}{\del \xi}+\frac{\rho R}{\kappa_1}\bigg(r^{1/2}\sin\pa{\frac{T_2-\xi}{\kappa_1}}-\frac{R}{2}\sin\pa{\frac{2(T_2-\xi)}{\kappa_1}}\bigg)\frac{\del T_2}{\del r}\\
&+\frac{\rho R}{2}\bigg(\frac{R}{2r}\pa{\cos\pa{\frac{2(T_2-\xi)}{\kappa_1}}-1}-\frac{\kappa_2}{\kappa_1 r^{1/2}}\sin\pa{\frac{T_2-\xi}{\kappa_1}}\bigg)\frac{\del T_2}{\del \xi}\\
&-\frac{\rho R}{2 \kappa_1r^{1/2}}\pa{\kappa_1 e^{T_2/\kappa_2}h_2(R^2)\cos\pa{\frac{T_2-\xi}{\kappa_1}}+\pa{\kappa_1 e^{T_2/\kappa_2}h_1(R^2)-\kappa_2}\sin\pa{\frac{T_2-\xi}{\kappa_1}}}
\end{aleq}%
\begin{aleq}\label{eq:dSxi}
\frac{\del S}{\del \xi}=&\frac{\rho r}{\kappa_1}\pa{R\sin\pa{\frac{2(T_2-\xi)}{\kappa_1}}-2r^{1/2}\sin\pa{\frac{T_2-\xi}{\kappa_1}}}\frac{\del R}{\del r}\\
&+\frac{\rho}{2}\bigg(R\pa{1-\cos\pa{\frac{2(T_2-\xi)}{\kappa_1}}}+\frac{r^{1/2}}{\kappa_1}\sin\pa{\frac{T_2-\xi}{\kappa_1}}\bigg)\frac{\del R}{\del\xi}\\
&+\frac{2r}{\kappa_1}\sin\pa{\frac{2(\kappa_1 T_1-\xi)}{\kappa_1}}\frac{\del T_1}{\del r}-\cos\pa{\frac{2(\kappa_1 T_1-\xi)}{\kappa_1}}\frac{\del T_1}{\del\xi}\\
&+\frac{\rho r R}{\kappa_1^2}\bigg(-2r^{1/2}\cos\pa{\frac{T_2-\xi}{\kappa_1}}+R\pa{1+\cos\pa{\frac{2(T_2-\xi)}{\kappa_1}}}\bigg)\frac{\del T_2}{\del r}\\
&+\frac{\rho R}{2\kappa_1^2}\pa{2\kappa_2 r^{1/2}\cos\pa{\frac{T_2-\xi}{\kappa_1}}+\kappa_1R\sin\pa{\frac{2(T_2-\xi)}{\kappa_1}}}\frac{\del T_2}{\del\xi}\\
&+\frac{\rho r^{1/2}R}{\kappa_1^2}\pa{\kappa_1 e^{T_2/\kappa_2}h_1(R^2)-\kappa_2}\cos\pa{\frac{T_2-\xi}{\kappa_1}}-\frac{\rho r^{1/2}R}{\kappa_1}e^{T_2/\kappa_2}h_2(R^2)\sin\pa{\frac{T_2-\xi}{\kappa_1}}
\end{aleq}%
\begin{aleq}\label{eq:red3}
&2\kappa_1 r\frac{\del R}{\del r}\cos\pa{\frac{T_2-2\kappa_1 T_1+\xi}{\kappa_1}}-\kappa_1^2\frac{\del R}{\del \xi}\sin\pa{\frac{T_2-2\kappa_1 T_1+\xi}{\kappa_1}}\\
&-2 r R\sin\pa{\frac{T_2-2\kappa_1 T_1+\xi}{\kappa_1}}\frac{\del T_2}{\del r}+\kappa_1 R\frac{\del T_2}{\del \xi}\cos\pa{\frac{T_2-2\kappa_1 T_1+\xi}{\kappa_1}}=0
\end{aleq}%
\begin{aleq}\label{eq:red4}
&2\kappa_1 r\frac{\del R}{\del r}\cos\pa{\frac{T_2-\xi}{\kappa_1}}+\kappa_1^2\frac{\del R}{\del \xi}\sin\pa{\frac{T_2-\xi}{\kappa_1}}\\
&-2rR\frac{\del T_2}{\del r}\sin\pa{\frac{T_2-\xi}{\kappa_1}}+\kappa_1 R\frac{\del T_2}{\del \xi}\cos\pa{\frac{T_2-\xi}{\kappa_1}}=0.
\end{aleq}%
Is should be noted that in order to obtain the reduced equations (\ref{eq:dSr}) and (\ref{eq:dSxi}) we must solve the equations resulting from the substitution of the Ansatz (\ref{eq:rf:2}) in (\ref{eq:1}.a) and (\ref{eq:1}.b) for $\del S/\del r$ and $\del S/\del\xi$. The reduced system consisting of equations (\ref{eq:dSr}), (\ref{eq:dSxi}), (\ref{eq:red3}) and (\ref{eq:red4}) is very complicated to solve in general. However, as an example, we find a particular solution by making the assumption that
\begin{eqe}\label{eq:rf:3}
R(r,\xi)=R(r),\qquad T_2(r,\xi)=T_2(\xi).
\end{eqe}%
In this case, equations (\ref{eq:red3}) and (\ref{eq:red4}) reduce to
\begin{gaeq}\label{eq:rf:4}
\kappa_1\cos\pa{\frac{T_2(\xi)-\xi}{\kappa_1}}\pa{2r\frac{d R(r)}{d r}+\frac{d T_2(\xi)}{d\xi}R(r)}=0,\\
\kappa_1\cos\pa{\frac{2\kappa_1T_1(r,\xi)-T_2(\xi)-\xi}{\kappa_1}}\pa{2 r\frac{dR(r)}{dr}-R(r)\frac{dT_2(\xi)}{d\xi}}=0.
\end{gaeq}%
One possible solution of the system (\ref{eq:rf:4}) is
\begin{eqe}\label{eq:rf:5}
R(r)=r^{1/2},\qquad T_2(\xi)=\xi+\kappa_1 \pi/2.
\end{eqe}%
The substitution of (\ref{eq:rf:5}) into the reduced equations (\ref{eq:dSr}) and (\ref{eq:dSxi}) results in the following system
\begin{aleq}\label{eq:rf:6}
\frac{\del S}{\del r}=&\frac{\del T_1}{\del r}\cos\pa{\frac{2}{\kappa}(\kappa_1T_1(\xi)-\xi)}+\frac{\kappa_1}{2r}\frac{\del T_1}{\del \xi}\sin\pa{\frac{2}{\kappa_1}(\kappa_1T_1(\xi)-\xi)}\\
&-2^{-1}\rho\pa{1+h_1(r)\exp\pa{\kappa_2^{-1}(\xi+\kappa_1\pi/2)}},\\
\frac{\del S}{\del \xi}=&\frac{2r}{\kappa_1}\frac{\del T_1}{\del r}\sin\pa{\frac{2}{\kappa_1}(\kappa_1T_1(\xi)-\xi)}-\frac{\del T_1(\xi)}{\del \xi}\cos\pa{\frac{2}{\kappa_1}(\kappa_1T_1(\xi)-\xi)}\\
&-\rho\kappa_1^{-1}\pa{1+h_2(r)\exp\pa{\kappa_2^{-1}(\xi+\kappa_1\pi/2)}},
\end{aleq}%
where the functions $h_1(r)$ and $h_2(r)$ are the same functions that define the force (\ref{eq:rf:1}). If we assume that
\begin{eqe}\label{eq:rf:7}
h_1(r)=2\frac{\kappa_2}{\kappa_1}\pa{h_2(r)+rh_2'(r)},
\end{eqe}
then the equations (\ref{eq:rf:6}) are compatible only if
\begin{eqe}\label{eq:rf:8}
T_1(r,\xi)=\frac{\xi}{\kappa_1}+\frac{\pi}{2}+\frac{1}{2}\arccos\pa{\frac{\rho}{2}r-\frac{a_2}{r}-a_3},
\end{eqe}%
where $a_2$ and $a_3$ are real constants. In this case, equations (\ref{eq:rf:6}) can be integrated and the obtained solution is
\begin{aleq}\label{defS}
S(r,\xi)=&-\frac{\rho r \xi}{\kappa_1}+\frac{r}{\kappa_1}\pa{\frac{1}{2\rho}-\frac{a_2}{r^2}}\xi -\frac{\xi}{\kappa_1}\pa{\frac{-\rho r}{2}+\frac{a_2}{r}+a_3}-\frac{\rho r}{2}\\
&+\frac{1}{2}\pa{1-\pa{\frac{\rho r}{2}-\frac{a_2}{r}-a_3}^2}^{1/2}+\frac{1}{2}\int r^{-1} \pa{1-\pa{\frac{\rho r}{2}-\frac{a_2}{r}-a_3}^2}^{1/2} dr\\
&-\frac{\kappa_2}{\kappa_1} \rho r \exp\pa{\kappa_2^{-1}\pa{\xi+\kappa_1 \pi/2}} h_2(r)+a_1.
\end{aleq}%
Finally, replacing the functions $R$, $T_1$, $T_2$ and $S$ defined in (\ref{eq:rf:5}), (\ref{eq:rf:8}) and (\ref{defS}) into the Ansatz (\ref{eq:rf:2}), we have the explicit solution
\begin{aleq}\label{eq:rf:9}
u=&-\frac{y}{t},\quad v=\frac{x}{t},\quad \theta=\frac{\pi}{2}+\arctan\pa{\frac{y}{x}}-\frac{1}{2}\arccos\pa{\frac{\rho}{2}\frac{x^2+y^2}{t^2}-\frac{a_2t^2}{x^2+y^2}-a_3},\\
\sigma=&-\frac{\rho\kappa_2}{\kappa_1}\frac{x^2+y^2}{t}h_2\pa{\frac{x^2+y^2}{t^2}}\exp\pa{\frac{\kappa_1}{\kappa_2}\pa{\arctan\pa{\frac{y}{x}}+\frac{\pi}{2}}}-\frac{\rho}{2}\frac{x^2+y2}{t^2}\\
&+\int^{r=\frac{x^2+y^2}{t^2}}\frac{1}{2r}\pa{1-\pa{\frac{\rho r}{2}-\frac{a_2}{r}-a_3}^2}^{1/2}dr+a_1\\
&+\frac{1}{2}\pa{1-\pa{\frac{\rho}{2}\frac{x^2+y^2}{t^2}-\frac{a_2t^2}{x^2+y^2}-a_3}^2}^{1/2}-a_3\pa{\frac{\kappa_2}{\kappa_1}\ln(t)+\arctan\pa{\frac{y}{x}}},
\end{aleq}%
where $a_1,a_2,a_3$ are integration constants. In (\ref{eq:rf:9}), the quadrature that appears in the formula for $\sigma$ can be computed in order to obtain an explicit expression in terms of hypergeometric functions. However, this expression is involved, so the author has prefered to give the more compact form of the quadrature.  In view of relation (\ref{eq:rf:7}) and replacing the explicit expression for the component of the velocity into the force (\ref{eq:fr:1}), we find the following explicit form for the force
\begin{aleq*}
F_1=&\left[\frac{2\kappa_2x-\kappa_1y}{\kappa_1t}h_2\pa{\frac{x^2+y^2}{t^2}}+\frac{2\kappa_2 x(x^2+y^2)}{\kappa_1 t^3}h_2'\pa{\frac{x^2+y^2}{t^2}}\right]\exp\pa{\frac{\kappa_1}{\kappa_2}\pa{\arctan\pa{\frac{y}{x}}+\frac{\pi}{2}}},\\
F_2=&\left[\frac{2\kappa_2y+\kappa_1x}{\kappa_1 t}h_2\pa{\frac{x^2+y^2}{t^2}}+\frac{2\kappa_2 y(x^2+y^2)}{\kappa_1 t^3} \right] \exp\pa{\frac{\kappa_1}{\kappa_2}\pa{\arctan\pa{\frac{y}{x}}+\frac{\pi}{2}}},
\end{aleq*}%
where $h_2$ is an arbitrary function defining the force.
\section{Concluding remarks and future outlook}\label{sec:con}
The objective of this paper was to study the system (\ref{eq:1})
describing the planar flow of an ideal plastic material in the
non-stationary case in order to obtain explicit solutions. The
symmetry group of the system depends on the components $F_1$ and
$F_2$ of the force involved in the system. The first stage was to
determine the symmetry group corresponding to each specific
investigated force. This investigation was carried out in Section
\ref{sec:2} where several different types of force were given together
which the associated symmetry generators. In many cases, the forces
depend on the components $u$ and $v$ of the velocity and include
arbitrary functions of one or two variables. For a given force (see
for example equation (\ref{eq:fr:1})) some symmetry generators
contain the parameters $\kappa_i$ of this force (see the generators
(\ref{eq:fr:2})). This corresponds to the fact that the symmetry
generator adapts itself to the force (through the value of the force
parameters). It should be noted that the monogenic forces of form
(\ref{eq:p:4}) are particularly interesting since the corresponding
symmetry group is of the highest dimension. In fact, some of the
generators (\ref{eq:8}) are expressible in terms of arbitrary
functions of time $\tau_2(t)$, $\tau_3(t)$ and $s(t)$, which makes
the group infinite dimensional. The subalgebra spanned by the eight
generators (\ref{eq:9}), corresponding to the requirement that
$\tau_2(t)$ and $\tau_3(t)$ be linear functions and that $s(t)$ be
constant, was classified into conjugacy classes under the action of
its internal automorphism group. This classification guarantees that
symmetry reductions corresponding to two different subalgebras
belonging to different conjugacy classes are not equivalent. By this
we mean that invariant solutions determined from two distinct
conjugacy classes cannot be obtained from each other through a group
transformation. This classification is summarized in Tables
\ref{tab:1}, \ref{tab:2}, and equations (\ref{eq:SC:30}) to (\ref{eq:classResult}) for subalgebras of
dimension 1 and 2. It can be used to carry out
symmetry reduction systematically in the case of monogenic forces.
Such a systematic procedure is not the objective of the present
paper but it is expected in a future work. However, in Section 3, we
use the conjugacy class represented by the subalgebra $\mathcal{L}_{2,1}=\ac{D,L}$ of table
\ref{tab:1} as an example of an invariant solution that can be
obtained in the case of monogenic forces. For a force of type
(\ref{eq:fr:1}), a second example of solution is given for which the
parameters $\kappa_1$ and $\kappa_2$ of the force appear in the
solution.
\paragraph{}As mentioned above a systematic use of the SRM on the classification given in Section 3, is a natural follow up of this work. The importance of such a study resides in the immediate applicability of the results, as was done in \cite{Lamothe:JMP:2012,Lamothe:JPA:2012} for the stationary case. New solutions of plasticity problems such as that given in system (\ref{eq:1}) are essential for the development and efficiency of certain industrial procedures such as sheet rolling and extrusion.
\section*{Acknowledgements}
The author is greatly indebted to professor A.M. Grundland (Centre de Recherche
Math\'ematiques, Universit\'e de Montr\'eal) for several valuable
and interesting discussions on the topic of this work. This work was supported by A.M. Grundland research grant from the Natural Sciences and Engineering Council of
Canada.
\section*{References}
\bibliographystyle{unsrt}

\begin{thebibliography}{10}

\bibitem{Kat:1}
Katchanov~L~1975
\newblock {\em \'El\'ements de la th\'eorie de la plasticit\'e}.
\newblock (\'Editions Mir, Moscou)

\bibitem{Hill}
Hill~R~1950
\newblock {\em The Mathematical Theory of plasticity}
\newblock (Oxford University press)

\bibitem{Chak}
Chakrabarty~J~2006
\newblock {\em Theory of Plasticity}.
\newblock (Elsevier)

\bibitem{Senashov:2007}
Senashov~S~I~and~Yakhno~A~2007
\newblock Reproduction of solutions of bidimensional ideal plasticity.
\newblock {\em International Journal of Non-Linear Mechanics}, 42:500--503

\bibitem{Senashov:2009}
Senashov~S~I,~Yakhno~A~and~Yakhno~L~2009
\newblock Deformation of characteristic curves of the plane ideal plasticity
  equations by point symmetries.
\newblock {\em Nonlinear analysis}, 2009.
\newblock doi:10.1016/j.na.2009.01.161.

\bibitem{Nadai:circularSol}
Nada~A.\"{\i}.~1924
\newblock \"{U}ber die gleit-und verweigungsfl\"{a}chen einiger gleinchgewichtszust\"{a}nde
  bildsamer massen und die nachspannungen bleibend verzenter k\"{o}rper.
\newblock {\em Z. Phys.}, 30(1):pp. 106--138

\bibitem{Prandtl:solPlas}
Prandtl~L.~1923
\newblock Anwendungsbeispeide zu einem henckychen satz \"{u}ber das plastiche
  gleichwitch.
\newblock {\em ZAMM}, 3(6):pp. 401--406

\bibitem{Czyz:1974}
Czyz~J~1974
\newblock Construction of a flow of an ideal plastic material in a die, on the
  basis of the method of {R}iemann invariants.
\newblock {\em Archives of Mechanics}, 26(4):589--616


\bibitem{Lamothe:JMP:2012}
Lamothe~V~2012
\newblock Symmetry group analysis of an ideal plastic flow.
\newblock {\em J. Math. Phys.}, 53, 033704

\bibitem{Lamothe:JPA:2012}
Lamothe~V~2012
\newblock Group analysis of an ideal plasticity model.
\newblock {\em J. Phys. A: Math. Theor.}, 45, 285203

\bibitem{PateraWinter:1}
Winternitz~P,~Patera~J~and~Zassenhaus~H 1975,
\newblock Continuous subgroups of the fundamental groups of physics. i. general
  method and the poincar\'e group.
\newblock {\em J. Math. Phys.}, 16:1597-1615

\bibitem{PateraWinter:2}
Sharp~R~T~Winternitz~P~Patera~J~and~Zassenhaus~H~1977,
\newblock Continous subgroup of the fundamental groups of physics. iii. the de
  sitter groups.
\newblock {\em J. Math. Phys.}, 18:2259

\bibitem{Ovsiannikov:Group_Analysis}
{Ovsiannikov~L~V~1982}
\newblock {\em Group Analysis of Differential Equations}.
\newblock (New-York, Academic Press)

\bibitem{Olver:Application_of_Lie}
{Olver~P~J~1986}
\newblock {\em Applications of Lie Groups to Differential Equations}.
\newblock (New-York, Springer-Verlag)

\bibitem{GrundlandLamothe:2013}
Grundland~A~M~and~Lamothe~V~2013
\newblock Multimode solutions of firts-order quasilinear systems obtained from Riemann invariants. Part I.
\newblock (submitted 2013)

\bibitem{PW:1}
Winternitz~P~1993
\newblock Lie groups and solutions of nonlinear partial differential equations.
\newblock Number CRM-1841, Centre de Recherches Math\'ematiques, Universit\'e de
  Montr\'eal

\bibitem{PateraWinter:1977}
Patera~J~and~Winternitz~P 1977
\newblock Subalgebras of real three- and four-dimensional Lie algebras.
\newblock {\em J. Math. Phys.}, 18:1449


\end{thebibliography}

\end{document}